\documentclass[nopacs,nokeys]{revtex4}
\usepackage{amsfonts}
\usepackage{amsmath}
\usepackage{amssymb}
\usepackage{bbm}
\usepackage[utf8]{inputenc}
\usepackage{tensor}
\usepackage{slashed}
\usepackage[centertableaux ]{ytableau}
\usepackage{hyperref}
\usepackage{natbib}
\usepackage{enumerate}
\usepackage{braket,mleftright}
\usepackage{graphicx}
\usepackage[caption=false]{subfig}
\usepackage{subfig}
\usepackage{cleveref}

\begin{document}

\title{Kinetic mixing, custodial symmetry, $Z$,  $Z^{\prime}$ interactions and $Z^{\prime}$ production in hadron colliders.}
 \author{ M. Napsuciale$^{(1)}$, S. Rodr\'{\i}guez$^{(2)}$,   H. Hern\'{a}ndez-Arellano$^{(1)}$ }
\address{$^{(1)}$Departamento de F\'{i}sica, Universidad de Guanajuato, Lomas del Campestre 103, Fraccionamiento
Lomas del Campestre, Le\'on, Guanajuato, M\'exico, 37150.}
\address{$^{(2)}$Facultad de Ciencias F\'isico-Matem\'aticas,
  Universidad Aut\'onoma de Coahuila, Edificio A, Unidad
  Camporredondo, 25000, Saltillo, Coahuila, M\'exico.}

\begin{abstract}
In this work we study the interactions of the $Z$ and $Z^{\prime}$ generated by kinetic mixing in a class of theories for physics 
beyond the standard model motivated by the dark matter problem, containing a spontaneously broken extra $U(1)_{d}$ 
gauge factor group in a hidden scenario and a Higgs sector respecting custodial symmetry. It is shown that custodial symmetry 
allows us to write the  $Z$ and $Z^{\prime}$ couplings 
to standard model fermions in terms of the measured values of $\alpha$, $G_{F}$, $M_{Z}$ and $M_{Z^{\prime}}$.
Working at the loop level, we calculate the ratio $\rho_{0}=M^{2}_{W}/\hat{c}^{2}_{Z}M^{2}_{Z}\hat{\rho}$ used in the 
fit to electroweak precision data (EWPD) and use its value to estimate possible effects of kinetic mixing at the electroweak scale.
For the $Z$ sector, we calculate the oblique parameters  $S$ and $T$, finding that for $M_{Z^{\prime}}\geq M_{Z}$ our results are in 
agreement with the values of the oblique parameters extracted from the global fit to EWPD at $1\sigma$ level. 
As to the $Z^{\prime}$ sector, we calculate the $Z^{\prime}$ contributions to charged lepton pair production at the Large Hadron Collider
in the well motivated case of dark matter entering 
particle physics as the matter fields of the $U(1)_{d}$ gauge symmetry with perturbative couplings at the electroweak scale, finding 
that data reported by the Compact Muon Selenoid Collaboration impose a lower limit $M_{Z^{\prime}}\gtrsim 5.0 ~TeV$. 
\end{abstract}
\maketitle

\section{Introduction}

Additional $U(1)$ gauge symmetries at low energies are predicted in a variety of models for physics beyond the standard model (SM). 
Grand unified theories with gauge groups of rank higher than the rank of the SM gauge group, yield naturally $U(1)$ factor groups  
\cite{Hewett:1988xc}, \cite{Langacker:2008yv}, and there exists a classification of these possibilities which have distinctive signatures 
for the low energy effects of a new massive physical neutral gauge boson, usually named $Z^{\prime}_{\mu}$ 
\cite{Holdom:1985ag},\cite{Dienes:1996zr},\cite{Babu:1997st}. 
One of the most important effects at low energies of this 
new physics, is the generation of a kinetic mixing between the $U(1)_{Y}$  and the extra $U(1)$ gauge bosons. Indeed, the renormalization 
group flow for the coupling constants from high to low energies yields a dimension four operator $B^{\mu\nu}V_{\mu\nu}$ where  $V_{\mu\nu}$ 
denotes the stress tensor for the $U(1)$ gauge boson $V_{\mu}$, even if it vanishes at some high energy scale 
\cite{Holdom:1985ag},\cite{Babu:1996vt}. This term, modifies the expected behavior of the low energy theory based only on the 
charges of SM fermions in the considered ultraviolet completion.  

There are two main effects of having a non-vanishing kinetic mixing at the electroweak scale. First, it produces a non-canonical form of the 
kinetic terms which can be diagonalized by a $GL(2,{\mathbb R})$ transformation generating small couplings of the new 
canonical field to the hypercharge of SM fields \cite{Babu:1997st}. These new couplings triggered the interest in using kinetic 
mixing as an alternative to conventional mechanisms to connect the SM with one of the most challenging problems today in high energy physics, 
the mystery of the nature of dark matter  \cite{Baumgart:2009tn}, 
\cite{Cheung:2009qd}, \cite{Ibarra:2009bm},\cite{Hook:2010tw}, \cite{Chun:2010ve}, \cite{Mambrini:2010dq}, \cite{Mambrini:2011dw}, 
\cite{Brahmachari:2014aya}, \cite{Arguelles:2016ney}, \cite{Belanger:2017vpq}, \cite{Arcadi:2018tly}, \cite{Foot:2012ai},
\cite{Kamada:2018kmi}, \cite{Rizzo:2018ntg}, \cite{Rizzo:2018joy}, \cite{Rizzo:2018vlb},  \cite{Banerjee:2019asa},  \cite{Rueter:2019wdf},
\cite{Akerib:2019diq}, \cite{Lao:2020inc}, \cite{Gehrlein:2019iwl}, \cite{Kribs:2020vyk}, \cite{Binh:2020xtf},\cite{Barnes:2020vsc}.
Second,  in the presence of kinetic mixing, new mass terms are generated by the spontaneous breaking of gauge symmetries.

In the standard model, the Higgs sector has a global $SU(2)_{L}\otimes SU(2)_{R}$ symmetry which under spontaneous
symmetry breaking (SSB) breaks down to its diagonal subgroup $SU(2)_{V}$. The $SU(2)_{L}$ gauge symmetry generators $\mathbf{T}$ 
transform as a triplet \cite{Weinberg:1975gm},\cite{Susskind:1978ms},\cite{Sikivie:1980hm} under this residual symmetry, named custodial 
symmetry in the literature \cite{Sikivie:1980hm}. This global symmetry requires the corresponding gauge bosons $\mathbf{W}_{\mu}$ 
to have a common mass and protects this property against radiative corrections. 
However, in the SM the $W^{3}_{\mu}$ component mixes with the gauge boson $B_{\mu}$ of the 
$U(1)_{Y}$ gauge symmetry to produce the massive $Z_{\mu}$ and the massless photon, such that the $Z_{\mu}$ mass is related to the
$W^{\pm}_{\mu}$ mass at tree level as $M_{W}=M_{Z} \cos\theta_{w}$. Radiative corrections involving $U(1)_{Y}$ charged operators yield 
small corrections to this mass relation which turn out to be of the order of one percent \cite{Zyla:2020zbs}. 

Physics beyond the standard model may modify this picture. New Higgs fields transforming in higher $SU(2)_{L}$ 
representations break custodial symmetry and modify the custodial symmetry relations \cite{Langacker:1991pg}. The value of the ratio 
$M^{2}_{W}/M^{2}_{Z}\cos^{2}\theta_{w}$ extracted from the global fit to electroweak precision data \cite{Zyla:2020zbs} 
puts stringent constraints on the possibility of custodial symmetry violating Higgs structures.

The relation $M_{W}=M_{Z} \cos\theta_{w}$ can also be modified by physics beyond the standard model with extra $U(1)$ gauge 
symmetries even if the Higgs sector respects custodial symmetry. In this case, the modification enters through the mixing of $W^{3}$ with 
more than one $U(1)$ gauge bosons. The stringent constraints on the photon mass \cite{Zyla:2020zbs} requires this mixing to 
preserve the unbroken nature of the electromagnetic $U(1)_{em}$ group generated by $Q=T^{3} + Y/2$, in which case, 
the custodial symmetry relation holds for a $\tilde{Z}_{\mu}$ field which is however not diagonal. The diagonalization procedure modifies 
the custodial symmetry relation for the physical field $Z_{\mu}$ and generates couplings of the physical extra gauge boson 
$Z^{\prime}_{\mu}$ to SM fields. The physical 
outcome of this scenario depends on details such as possible charges of SM fields of the new $U(1)$ gauge symmetries and if the new 
fields carry SM charges. Under SSB these charges may generate new mass terms yielding a rich scenario which is 
however constrained by electroweak precision data. 

The present work is motivated by the dark matter connection, where fields in the ultraviolet completion belong to the dark matter sector.
In this case, the coupling of SM fields with the dark fields must be very small and the simplest realization is to have separate 
SM and dark worlds, connected only by the kinetic mixing of $U(1)_{Y}$ and the $U(1)$ factor subgroup of the gauge symmetry of 
the dark sector. This hidden dark matter scenario is an alternative realization of the Weakly Interactive Massive Particle (WIMP) idea
where the small couplings arises from perturbative gauge couplings and massive mediator between dark and SM fields. The possibility of a
kinetic mixing connection between dark and standard model sectors has been previously considered in the literature, but we will work 
out here the consequences of a physically well motivated constraint for the ultraviolet completion: custodial symmetry. 

It was recently shown that if custodial symmetry is respected by the extended Higgs sector, the fact that the mass term of the $W^{3}_{\mu}$ 
is related by this symmetry to the mass term of the $W^{\pm}_{\mu}$, can be used to write the mixing parameters entirely in terms of SM observables and the mass of the $Z^{\prime}_{\mu}$ \cite{Napsuciale:2020kai} . This procedure has the advantage that we can use 
precision data of the SM to constrain directly the mass of the new gauge boson instead of the mixing parameters as is conventionally 
done. Using the results of the global fit to EWPD for the ratio $\rho_{0}=M^{2}_{W}/M^{2}_{Z}\hat{c}^{2}_{Z}\hat{\rho}$ at $1\sigma$ level
\cite{Zyla:2020zbs}, we did show in \cite{Napsuciale:2020kai} that, in the considered framework, mixing relations straightforwardly yield 
the lower bound $M_{Z^{\prime}}>M_{Z}$.

In this framework, the couplings of the $Z_{\mu}$ and $Z^{\prime}_{\mu}$ bosons in the extended theory can also be written in terms 
of measured SM data and the mass of the $Z^{\prime}_{\mu}$ boson. In the present work we do this rewriting and study the implications 
for the physics of the $Z_{\mu}$ boson and possible effects at the electroweak scale of the existence of a $Z^{\prime}_{\mu}$ boson. 
As to the $Z_{\mu}$ 
boson, effects at the electroweak scale of physics beyond the SM can in general be encoded in the oblique parameters $S$, $T$ and 
$U$ \cite{Lynn:1985fg},\cite{Kennedy:1988sn},\cite{Kuroda:1990wn}, \cite{Peskin:1990zt}, \cite{Peskin:1991sw}, thus we focus on the 
calculation of these parameters. Concerning effects of the $Z^{\prime}_{\mu}$ at low energies, we calculate its contribution to the  
production of a lepton pair in hadron colliders and compare our results with experimental data obtained by 
the Compact Muon Selenoid (CMS) collaboration at the Large Hadron Collider (LHC) \cite{Sirunyan:2018exx},\cite{CMS:2021ctt} .

Our work is organized as follows. In the next section we rewrite the couplings of the neutral gauge bosons 
in theories for physics beyond the SM with an extra $U(1)$ factor group, kinetic mixing and a Higgs sector respecting custodial 
symmetry. Section III is devoted to the calculation of the oblique parameters and a comparison with results of the global fit to EWPD. 
In section IV we calculate the induced $Z^{\prime}_{\mu}$ couplings to SM fermions, work out the $Z^{\prime}_{\mu}$ contributions 
to lepton pair production at the LHC and compare with CMS results. Our conclusions are given in Section V.    

\section{Additional $U(1)_{d}$, kinetic mixing and custodial symmetry}
Let us consider an extension of the SM to a group $G$ with a spontaneously broken factor Abelian gauge symmetry which we will denote 
as $U(1)_{d}$ in the following. At low energies the theory has the $G_{SM}\otimes U(1)_{d}$ gauge symmetry and the
invariant  Lagrangian including all dimension-four terms is given by 
\begin{equation}
\mathcal{L}=\mathcal{L}_{SM}(\tilde W^{a},\tilde B,\tilde\phi) + \mathcal{L}_{V}(V, \Phi) - \frac{ \sin\chi }{2}V^{\mu\nu}\tilde B_{\mu\nu} 
- 2 \kappa \Phi^{*}\Phi \tilde\phi^{\dagger}\tilde\phi ,
\label{Lag}
\end{equation}
where we use $\tilde{f}$ to distinguish the SM fields $f$ in the extended theory, $\tilde B^{\mu\nu}$ stands for the  $U(1)_{Y}$ 
strength tensor, $V^{\mu\nu}$ denotes the strength tensor for the extra gauge boson $V_{\mu}$ of $U(1)_{d}$ , and the complex 
Higgs field which spontaneously breaks this symmetry is denoted as $\Phi$. The explicit form of 
$ \mathcal{L}_{V}(V, \Phi)$ depends on our choice of the ultraviolet completing theory, but results in this paper are 
independent of this choice, except for the structure of the Higgs sector, which upon spontaneous symmetry breaking 
must respect custodial symmetry.

The kinetic terms for the gauge bosons in the Lagrangian in Eq. (\ref{Lag}) are
\begin{equation}
\mathcal{L}^{K}_{gauge}=-\frac{1}{4}( \tilde W^{a\mu\nu}\tilde W^{a}_{\mu\nu} + \tilde B^{\mu\nu}\tilde B_{\mu\nu}
+ V^{\mu\nu}V_{\mu\nu} + 2 \sin\chi V^{\mu\nu}\tilde B_{\mu\nu} ), 
\label{Klag}
\end{equation}
and contain a kinetic mixing term of the gauge bosons of $U(1)_{Y}$ and $U(1)_{d}$ factor groups. This term makes the 
kinetic Lagrangian not canonical and we must perform the following $GL(2,\mathbb{R})$ transformation on the gauge bosons to 
obtain properly normalized kinetic terms \cite{ Babu:1997st}
\begin{align}
\tilde B_{\mu\nu}&=\bar{B}_{\mu\nu}-\tan\chi \bar{V}_{\mu\nu},  
\qquad V_{\mu\nu}=\sec\chi \bar{V}_{\mu\nu}.
\end{align}
After this transformation, the kinetic terms gets the canonical form
\begin{equation}
\mathcal{L}^{K}_{gauge}=-\frac{1}{4}( \tilde W^{a\mu\nu} \tilde W^{a}_{\mu\nu}+ \bar{B}^{\mu\nu}\bar{B}_{\mu\nu}
+\bar{V}^{\mu\nu}\bar{V}_{\mu\nu}),
\label{Klagd}
\end{equation}
but it induces a coupling of the $U(1)_{d}$ gauge boson with the SM fields. Indeed, after this transformation 
the $SU(2)_{L}\otimes U(1)_Y\otimes U(1)_{d}$ covariant derivative  reads 
\begin{equation}
D^{\mu}=\partial^{\mu} + i \tilde g T^{a}\tilde W^{a\mu}  + i\tilde g_{Y} \frac{Y}{2} \bar{B}^{\mu} 
+ i (g_{d}  \sec\chi \frac{Q_{d}}{2}-\tilde g_{Y} \tan\chi \frac{Y}{2}) \bar{V}^{\mu},
\end{equation}
where $ Q_{d}/2$ denotes the generator of $U(1)_{d}$, $g_{d}$ is the corresponding coupling constant and we use the same 
"tilde" notation for the SM electroweak gauge couplings, $\tilde{g}, \tilde{g}_{Y}$, in the extended theory.

The effect of the kinetic mixing propagates and reaches mass terms generated by the Higgs mechanism causing a mixing 
of the SM neutral bosons with the $\bar V_{\mu}$ gauge boson to produce the physical $A_{\mu}$, $Z_{\mu}$ and a new physical 
boson denoted by $Z^{\prime}_{\mu}$. Indeed, if we want to keep the $U(1)_{em}$ unbroken with a generator $Q=T_{3}+Y/2$, 
we need the $\bar{B}_{\mu}$ to be the SM hypercharge gauge boson which mixes with $\tilde W^{3}_{\mu}$ to produce the physical photon. 
Notice that the  $\bar{B}_{\mu}$ field has the same couplings to SM fields as the original $\tilde{B}_{\mu}$ field thus this is just a 
reinterpretation of the SM $U(1)_{Y}$ gauge boson. With the conventional weak rotation 
\begin{equation}
\begin{pmatrix}\bar{B}\\ \tilde W_{3} \end{pmatrix}
= \begin{pmatrix}\cos\tilde \theta_{w}&\ -\sin\tilde \theta_{w} \\ \sin\tilde \theta_{w} & \cos\tilde \theta_{w} \end{pmatrix} 
\begin{pmatrix} A\\ \tilde Z  \end{pmatrix}
\end{equation}
we obtain
\begin{align}
\tilde{g} T_{3}\tilde W_{3} + \tilde{g}_{Y} \frac{Y}{2} \bar{B} 
= e Q A  + \frac{\tilde{g}}{ \tilde{c}_{w}}  ( T_{3} - \tilde{s}^{2}_{w} Q  ) \tilde Z ,
\end{align}
with $e=\tilde{g} \tilde {s}_{w} =\tilde{g}_{Y} \tilde{c}_{w} $.
Hereafter, we will use the shorthand notation 
$\tilde{s}_{x} =\sin\tilde\theta_{x}$, $\tilde{c}_{x} =\cos\tilde\theta_{x}$ for the mixing angles.

The Lagrangian for the Higgs sector of the $G_{SM}\otimes U(1)_{d}$ gauge theory reads
\begin{equation}
{\cal L}_{Higgs}=(D^{\mu}\tilde\phi)^{\dagger}D_{\mu}\tilde\phi  + (D^{\mu}\Phi)^{*}D_{\mu}\Phi - V(\tilde\phi,\Phi) ,
\label{Higgslag}
\end{equation}
with the following Higgs potential
\begin{align}
V(\tilde\phi,\Phi) &=\tilde\mu^{2}\tilde\phi^{\dagger}\tilde\phi + \tilde\lambda (\tilde\phi^{\dagger}\tilde\phi)^{2} 
+\mu^{2}_{d}\Phi^{*}\Phi + \lambda_{d} (\Phi^{*}\Phi)^{2} 
+2 \kappa \Phi^{*}\Phi \tilde\phi^{\dagger}\tilde\phi .
\label{Higgspot}
\end{align}

We are interested in the effects of kinetic mixing here, thus we will assume that the SM Higgs $\tilde \phi$ is a singlet under $U(1)_{d}$ 
 and the new Higgs field $\Phi$ is a singlet under the SM group. In the unitary 
gauge, Eq.(\ref{Higgslag}) yields the following gauge bosons mass terms
\begin{equation}
{\cal L}_{mass}=M^{2}_{\tilde W} \tilde W^{+\mu}\tilde W^{-}_{\mu}
+ \frac{1}{2}\left[ M^{2}_{\tilde Z} \tilde Z^{2} +2\Delta~ \bar{V} ^{\mu} \tilde Z_{\mu}
 + M^{2}_{\bar{V}} \bar{V}^{2} \right] ,
\end{equation} 
with
\begin{equation}
M^{2}_{\tilde W}= \frac{\tilde g^{2}\tilde v^{2}}{4}, \qquad 
M^{2}_{\tilde Z}=\frac{M^{2}_{\tilde W}}{\tilde c^{2}_{w}}, \qquad
\Delta= \frac{M^{2}_{\tilde W}}{\tilde c^{2}_{w}} \tilde s_{w} \tan\chi, \qquad
 M^{2}_{\bar{V}}=M^{2}_{\tilde W} \tan^{2}\tilde\theta_{w} \tan^{2}\chi + g^{2}_{d} v_{d}^{2} \sec^{2}\chi   .
 \label{Delta}
\end{equation} 
The photon is massless, the SM field $\tilde Z$ has the expected mass value from custodial symmetry but a mixing with 
the $\bar V$ field has been generated by the kinetic mixing. The neutral gauge boson part of this Lagrangian can be 
diagonalized by the following rotation
\begin{equation}
\begin{pmatrix}\tilde Z\\ \bar{V} \end{pmatrix}= \begin{pmatrix}\cos\theta_{\zeta} &-\sin\theta_{\zeta}  \\ 
\sin\theta_{\zeta}  & \cos\theta_{\zeta}  \end{pmatrix} 
\begin{pmatrix}Z\\\ Z'  \end{pmatrix}.
\label{rot}
\end{equation}
After these transformations, the original gauge fields are related to the diagonal fields by the following matrix
\begin{align}
\begin{pmatrix}\tilde B \\ \tilde W_{3} \\ V\end{pmatrix}&= \begin{pmatrix} \tilde{c}_{w}, &  - \tilde{s}_{w} c_{\zeta}-\tan\chi s_{\zeta}, 
& \tilde{s}_{w} s_{\zeta}-\tan\chi c_{\zeta} \\
\tilde{s}_{w} &\tilde{c}_{w} c_{\zeta}&-\tilde{c}_{w} s_{\zeta}   \\ 
0&\sec\chi s_{\zeta} & \sec\chi c_{\zeta} \end{pmatrix} \begin{pmatrix}  A \\ Z\\ Z' \end{pmatrix} ,
\label{mixmat}
\end{align}
and in terms of the physical fields the covariant derivative reads
\begin{align}
D_{\mu}&=\partial_{\mu} + i \frac{\tilde g }{\sqrt{2}} (T^{+}\tilde W^{+}_{\mu}+T^{-}\tilde W^{-}_{\mu} )  + i  e Q  A_{\mu}  \nonumber \\
&+ i \left[\frac{\tilde g  c_{\zeta}}{ \tilde c_{w}} \left(  ( T_{3} - \tilde s^{2}_{w} Q  ) - \tilde s_{w} \tan\theta_{\zeta} \tan\chi \frac{Y}{2}\right) 
+ g_{d} s_{\zeta} \sec\chi \frac{Q_{d}}{2} \right]Z_{\mu} \nonumber \\
& - i \left[ \frac{\tilde g s_{\zeta} }{ \tilde c_{w}} \left(  ( T_{3} - \tilde s^{2}_{w} Q  ) +\frac { \tilde s_{w} \tan\chi}{\tan\theta_{\zeta}} \frac{Y}{2}  \right) 
- g_{d} c_{\zeta} \sec\chi \frac{Q_{d}}{2}\right] Z^{\prime}_{\mu} .
\label{covder}
\end{align}

The effects of kinetic mixing are conventionally analyzed comparing experimental data with predictions from the neutral currents 
arising from this covariant derivative. This comparison yields usually bounds of the possible values of the kinetic mixing parameter 
$\chi$, the mixing angle $\theta_{\zeta}$ or the mixing angle in the Higgs sector due to the $\kappa$ term in 
Eq. (\ref{Higgspot}), not shown here. 
In a recent work \cite{Napsuciale:2020kai} we pointed out that the tree level custodial symmetry protected relation 
$M^{2}_{\tilde W}=M^{2}_{\tilde Z} \cos^{2}\tilde \theta_{w}$, allows to write the matrix elements in Eq. (\ref{mixmat}) in terms of the 
measured weak angle, $M_{W}$, $M_{Z}$ and the unknown mass of the diagonal field $Z^{\prime}$. Aiming to use results of the 
fit to electroweak precision data which requires a loop level analysis, we argue that loop contributions are dominated by the 
effects of standard model particles, and the main effect of kinetic mixing in the mass Lagrangian can be taken at the tree level.  

Effects of new physics are considered in the global fit to the EWPD through the parameter \cite{Zyla:2020zbs}
\begin{equation}
\rho_{0}\equiv \frac{M^{2}_{W}}{\hat c^{2}_{Z}M^{2}_{Z}\hat{\rho}} ,
\label{rho0}
\end{equation}
where $\hat c_{Z}\equiv \cos \theta_{w} (M_{Z})$ is the Weinberg angle measured at the $Z$ pole and $M^{2}_{W}$, $M^{2}_{Z}$ 
denote the masses of the physical $Z$ and $W^{\pm}$ respectively. The quantity $\hat{\rho}$ in Eq. (\ref{rho0}) accounts for radiative 
corrections in the SM such that if there are no new physics contributions then $\rho_{0}=1$. Deviations from this value are 
necessarily due to physics beyond the SM. 

Radiative corrections to the ratio $M^{2}_{W}/\hat c^{2}_{Z}M^{2}_{Z}$ due to SM particles
are dominated by the top quark and the next to leading order effects are given by Higgs boson loops.
In the $\overline{MS}$ scheme, including all boson contributions yields \cite{Zyla:2020zbs} 
\begin{equation}
\hat{\rho}=1.01019\pm0.00009.
\label{rho0rad}
\end{equation}
The global fit to electroweak precision data yields  \cite{Zyla:2020zbs}
\begin{equation}
\rho_{0}=1.00038\pm 0.00020.
\label{rho0exp}
\end{equation}
Notice that $\rho_{0}>1$ at $1.9 \, \sigma$ ($94\%$ confidence level)  \cite{Zyla:2020zbs}. Although not 
conclusive, this result points to possible new physics contributions to the value of $\rho_{0}$ which we will assume to be dominated by the 
kinetic mixing effects. In the following we will explore the consequences of having $\rho_{0}\neq1$, aiming to constrain the possible values of
the $Z^{\prime}$ mass from available data. 

Comparison with electroweak precision data requires to calculate $\rho_{0}$ in the extended theory. Since in the hidden $U(1)_{d}$
extension of the standard model $M_{\tilde{W}}$ is the physical mass of the $W^{\pm}$ we have
\begin{equation}
\rho_{0}=\frac{M^{2}_{\tilde{W}}}{\hat{c}^{2}_{Z}M^{2}_{Z}\hat{\rho}}.
\end{equation}
Using custodial symmetry relations at the loop level it has been shown that the $\tilde{Z}-\bar{V}$ mixing angle in the 
$\overline{MS}$ scheme
is given by \cite{Napsuciale:2020kai}
 \begin{equation}
\hat{s}^{2}_{\zeta}=\frac{\sigma_{0} (\rho_{0}-1)(\rho_{0}\hat{c}^{2}_{Z} - \hat{s}^{2}_{Z} )}{(\rho_{0} - \hat{s}^{2}_{Z})(\rho_{0}-\sigma_{0}) }, 
\end{equation}
where $\hat{s}_{\zeta}=s_{\zeta}(M_{Z})$,  is the sine of $\tilde{Z}-\bar{V}$ mixing angle at the scale $\mu=M_{Z}$ and
\begin{equation}
\sigma_{0}\equiv \frac{M^{2}_{\tilde{W}}}{\hat{c}^{2}_{Z}M^{2}_{Z^{\prime}}\hat{\rho}}.
\label{rho0s2hats2}
 \end{equation}

Similarly, the kinetic mixing angle at the same scale, $\hat\chi=\chi(M_{Z})$, can be written in terms of the same physical 
quantities as 
\begin{equation}
\tan^{2}\hat\chi= \frac{(\rho_{0}-1) (\rho_{0} \hat{c}^{2}_{Z} - \hat{s}^{2}_{Z})}
{\rho^{2}_{0}\sigma_{0} \hat{s}^{2}_{Z} \hat{c}^{4}_{Z}} [\rho_{0}(1-\sigma_{0} \hat{c}^{2}_{Z}) - \hat{s}^{2}_{Z}].
\label{tchi2tz2}
\end{equation}

Using these relations, the covariant derivative in Eq. (\ref{covder}) can be rewritten in terms of the physical values of 
the effective couplings depending on $\alpha,G_{F}, M_{W}, M_{Z}$ and $M_{Z^{\prime}}$ as 
\begin{align}
D_{\mu}&=\partial_{\mu} + i \frac{e \sqrt{\rho_{0} }}{\sqrt{2}\hat{s}_{Z}} (T^{+}W^{+}_{\mu}+T^{-}W^{-}_{\mu} )  + i  e Q A_{\mu}  \nonumber \\
&+ i \left[\frac{e  }{\hat{s}_{Z}  \hat{c}_{Z}} \sqrt{\frac{\rho_0 - \hat{s}^{2}_{Z} -\rho_{0}\sigma_{0} \hat{c}^{2}_{Z}}
{\hat{c}^{2}_{Z}\rho_{0}(\rho_{0}-\sigma_{0})}}
\left( T_{3} - (1-\rho_{0} \hat{c}^{2}_{Z} )Q  \right)  + g_{d} \hat{s}_{\zeta} \sec\hat\chi \frac{Q_{d}}{2} \right]Z_{\mu} \nonumber \\
& - i \left[ \frac{e  }{\hat{s}_{Z} \hat{c}_{Z}} \sqrt{\frac{(\rho_0-1)(\rho_{0}\hat{c}^{2}_{Z}-\hat{s}^{2}_{Z} )}{\hat{c}^{2}_{Z} \sigma_{0}(\rho_{0}-\sigma_{0})}}
\left( T_{3} - (1-\sigma_{0} \hat{c}^{2}_{Z} )Q  \right)  - g_{d} \hat{c}_{\zeta} \sec\hat\chi \frac{Q_{d}}{2}\right] Z^{\prime}_{\mu},
\label{covder2}
\end{align}
where 
\begin{align}
\hat{c}^{2}_{\zeta}&=\frac{\rho_{0} (\rho_{0} - \hat{s}^{2}_{Z}-\rho_{0}\sigma_{0}\hat{c}^{2}_{Z} )}{(\rho_{0} 
- \hat{s}^{2}_{Z})(\rho_{0}-\sigma_{0}) } ,\\
\hat{s}_{\zeta} \sec\hat\chi &=\frac{1}{\rho_{0}\hat{s}_{Z}\hat{c}^{2}_{Z} }\left[ \frac{(\rho_{0}-1) (\rho_{0}\hat{c}^{2}_{Z} - \hat{s}^{2}_{Z} ) 
(\rho_{0}(\rho_{0}\hat{c}^{2}_{Z} -1)(1-\sigma_{0}\hat{c}^{2}_{Z} ) + \hat{s}^{2}_{Z} )}{\rho_{0}-\sigma_{0}}  \right]^{\frac{1}{2}} , \\
\hat{c}_{\zeta} \sec\hat\chi &= \frac{1}{\hat{s}_{Z}\hat{c}_{Z}}\left[ \frac{\rho_{0}- \hat{s}^{2}_{Z} -\rho_{0}\sigma_{0}\hat{c}^{2}_{Z} }{\rho_{0}-\sigma_{0}}
\left( 
\frac{(\rho_{0}-1)}{\sigma_{0}} \frac{(\rho_{0}(1-\sigma_{0})\hat{c}^{2}_{Z} - \hat{s}^{2}_{Z} )}{\rho_{0}\hat{c}^{2}_{Z} }  +\rho_{0}\hat{s}^{2}_{Z}  
\right)  
\right]^{\frac{1}{2}},
\label{czsecchi}
\end{align}
with $\hat{c}_{\zeta}=c_{\zeta}(M_{Z})$.

\section{ Effective $Z\bar{f}f$ interactions and oblique parameters}

The effective Lagrangian formulation for physics beyond the SM encode corrections to the SM Lagrangian in the oblique parameters 
$S$, $T$ and $U$. For the $Z\bar{f}f$ interaction the effective Lagrangian reads \cite{Babu:1997st},\cite{Holdom:1990xp},\cite{Golden:1990ig}, 
\cite{Altarelli:1990zd}, \cite{ Burgess:1993mg}
\begin{equation}
{\cal L}^{eff}_{Z\bar{f}f}=\frac{e}{2s_{W}c_{W}} \left(1+\frac{\alpha T}{2} \right) \sum_{f} \bar{f}\gamma^{\mu}
\left( T^{3}_{f_{L}} -2 s^{2}_{\ast} Q - T^{3}_{f_{L}} \gamma^{5} \right)f Z_{\mu} ,
\label{eftlag}
\end{equation}
where $s_{W}=\sin\theta_{W}$,  $c_{W}=\cos\theta_{W}$ with $\theta_{W}$ standing for the value of the Weinberg angle at the 
considered low energy scale and 
\begin{equation}
s^{2}_{\ast}= s^{2}_{W} + \frac{1}{c^{2}_{W}-s^{2}_{W}} \left( \frac{\alpha S}{4} -s^{2}_{W}c^{2}_{W} \alpha T \right).
\end{equation}

If the SM fermions do not carry $U(1)_{d}$ charge, the covariant derivative in Eq.(\ref{covder2}) yields the following $Z\bar{f}f$ 
Lagrangian at the scale $\mu=M_{Z}$
\begin{equation}
{\cal L}_{Z\bar{f}f}=\frac{e}{2\hat{s}_{Z} \hat{c}_{Z}} R \sum_{f} \bar{f}\gamma^{\mu}
\left[ T^{3}_{f_{L}} -2 (1-\rho_{0}\hat{c}^{2}_{Z}) Q - T^{3}_{f_{L}} \gamma^{5} \right] f~ Z_{\mu} ,
\label{lagZff}
\end{equation}
with
\begin{equation}
R= \sqrt{\frac{\rho_0 - \hat{s}^{2}_{Z} -\rho_{0}\sigma_{0}\hat{c}^{2}_{Z}}{\hat{c}^{2}_{Z}\rho_{0}(\rho_{0}-\sigma_{0})}} .
\end{equation}

A comparison of Lagrangians in Eqs. (\ref{eftlag},\ref{lagZff}) yields
\begin{align}
\alpha S&= 4 \hat{c}^{2}_{Z} \left[ (1-\rho_{0})(\hat{c}^{2}_{Z} - \hat{s}^{2}_{Z}) + 2 \hat{s}^{2}_{Z} (R-1) \right] ,\\
\alpha T&=2 (R-1) .
\end{align}
Notice that although $S$ and $T$ are functions of the $Z^{\prime}$ mass, there is always a linear relation among these parameters
\begin{equation}
T=\frac{1}{4\hat{s}^{2}_{Z}\hat{c}^{2}_{Z} }S + (\rho_{0}-1)\frac{\hat{c}^{2}_{Z} - \hat{s}^{2}_{Z} }{\alpha \hat{s}^{2}_{Z}} ,
\end{equation}
valid for all values of $M_{Z^{\prime}}$.
The values extracted from the fit to EWPD for the oblique parameters are \cite{Zyla:2020zbs}
\begin{equation}
S=-0.01\pm 0.10, \qquad T=0.03 \pm 0.12, \qquad U=0.02 \pm 0.11.
\label{STUexp}
\end{equation}

\begin{figure}[h]
\includegraphics[scale=0.4]{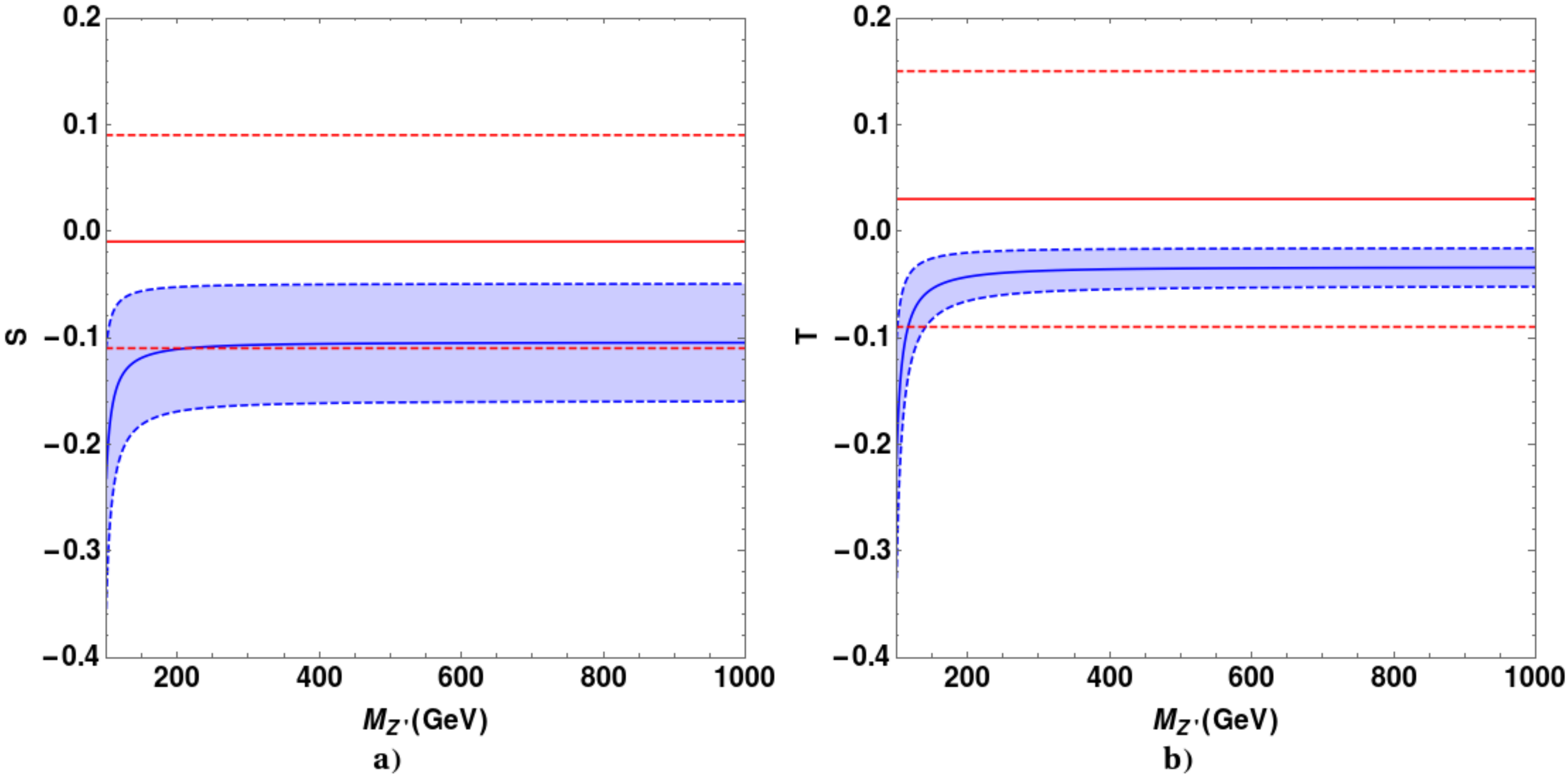}
\caption{Oblique parameters $S$ and $T$ as functions of $M_{Z^{\prime}}$. The solid blue lines corresponds to the predictions using the 
central value of $\rho_0$ and the shadow band to the $1\sigma$ region for $\rho_{0}$  in Eq. (\ref{rho0exp}). The red 
bands correspond to the $1\sigma$ region for $S$ and $T$ in Eq.(\ref{STUexp}). Solid lines correspond to the central values. }%
\label{STplot}%
\end{figure}

\begin{figure}[h]
\includegraphics[scale=0.4]{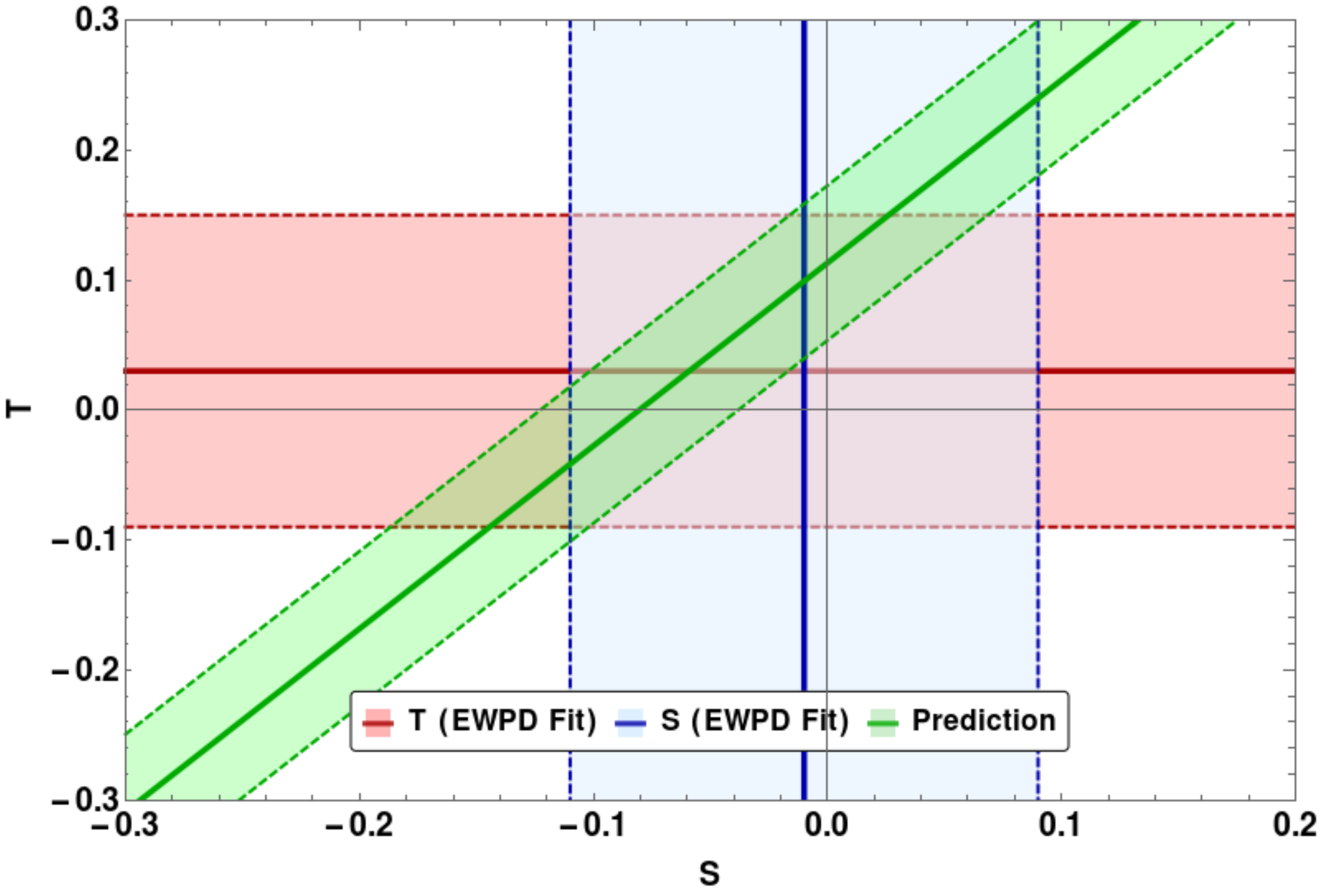}
\caption{Oblique parameter $S$ as a function of $T$. The green band corresponds to the predictions using the $1\sigma$ region for 
$\rho_{0}$ in Eq. (\ref{rho0exp}). The red and blue bands corresponds values of $T$ and $S$ respectively in Eq.(\ref{STUexp}) 
considering the $1\sigma$ region. Solid lines correspond to the central values.}%
\label{TvsSplot}%
\end{figure}

In Fig. (\ref{STplot}) we plot the predicted values of $S$ and $T$ as functions of the $Z^{\prime}$ mass, for the range 
of values for $\rho_{0}$ extracted from the fit to EWPD in Eq.(\ref{STUexp}). We also show in these plots the $1\sigma$ regions for 
$S$ and $T$ obtained in the fit. We notice first that $S$ and $T$ reach a saturation value for $M_{Z^{\prime}}\approx 250 ~GeV$ 
and are not sensitive to  the value of $M_{Z^{\prime}}$ beyond this point. The predicted values of $S$ and $T$ are consistent with 
results from the fit to EWPD for   $M_{Z^{\prime}}> M_{Z}$. The linear relation between $T$ and $S$ is shown in Fig.(\ref{TvsSplot}) together 
with the $1\sigma$ bands for the oblique parameters. The intersection of the three bands yields the values of $S$ and $T$ for 
which the predictions of the present formalism agrees with the fit to EWPD at $1\sigma$ level. 
\section{$Z^\prime$ contributions to charged lepton pair production at hadron colliders}
\subsection{General formalism}

Upper bounds have been obtained for the $Z^{\prime}$ contributions to charged lepton pair production at the 
Tevatron \cite{ABAZOV201188} and the LHC \cite{Sirunyan:2018exx},\cite{CMS:2021ctt}. The production 
cross section for a fermion pair in hadron colliders in the $Z^{\prime}$ pole region can be written in general as 
\cite{Carena:2004xs}, \cite{Accomando:2010fz}
\begin{equation}
\sigma_{\bar{f}f}= \int_{(M_{Z\prime}-\Delta)^{2}}^{(M_{Z\prime}+\Delta)^q{2}} \frac{d\sigma}{dM^{2}} (pp \to Z^{\prime} X \to \bar{f}f X)dM^{2}.
\end{equation}
In the narrow width approximation for the $Z^{\prime}$ it can be written as \cite{Carena:2004xs}
\begin{equation}
\sigma_{\bar{f}f}\approx \left( \frac{1}{3} \sum_{q=u,d} \frac{dL_{\bar{q}q}}{dM^{2}_{Z^{\prime}} } \hat\sigma (\bar{q}q \to Z^{\prime}) \right)
BR(Z^{\prime} \to \bar{f}f ),
\end{equation}
where $\frac{dL_{\bar{q}q}}{dM^{2}_{Z^{\prime}}}$ stands for the parton luminosities and the branching ratio for the $\bar{f}f$ channel 
is given by
\begin{equation}
BR(Z^{\prime} \to \bar{f}f )=\frac{\Gamma(Z^{\prime} \to \bar{f}f )}{\Gamma_{Z^{\prime} }},
\end{equation}
where $\Gamma_{Z^{\prime}}$ denotes the total $Z^{\prime}$ width. 

In general, the Lagrangian for the interaction of the $Z^{\prime}$ with standard model fermions can be written as  \cite{Accomando:2010fz}
 \begin{equation}
{\cal L}_{Z^{\prime}\bar{f}f}=g^{\prime}  Z^{\prime}_{\mu}  \sum_{f} \bar{f}\gamma^{\mu}
\left[ g^{f}_{V} - g^{f}_{A }\gamma^{5} \right] f ,
\label{lagzpgen}
\end{equation}
such that the peak cross section reads
\begin{equation}
\hat\sigma (\bar{q}q \to Z^{\prime}) = \frac{\pi g^{\prime 2}}{12} \left[ (g^{q}_{V})^{2} + (g^{q}_{A})^{2} \right],
\end{equation}
and the width in the $\bar{f}f$ channel is calculated as
\begin{equation}
\Gamma(Z^{\prime} \to \bar{f}f )= N_{c} \frac{g^{\prime 2}M_{Z^{\prime}}}{48\pi} \left[ (g^{f}_{V})^{2} + (g^{f}_{A})^{2} 
+ {\cal O}(\frac{m^{2}_{f}}{M^{2}_{Z^{\prime}}} ) \right],
\end{equation}
where $N_{c}=3$ for quarks and $N_{c}=1$ for leptons. Neglecting the SM fermion masses, if the couplings are generation independent, the total width to fermions is given by
\begin{equation}
\Gamma^{f}_{Z^{\prime}}  =  \frac{g^{\prime 2}M_{Z^{\prime}}}{48\pi} \left[ 9\left( (g^{u}_{V})^{2} + (g^{u}_{A})^{2} 
+ (g^{d}_{V})^{2} + (g^{d}_{A})^{2}\right) +3  \left( (g^{\nu}_{V})^{2} + (g^{\nu}_{A})^{2} + (g^{e}_{V})^{2} + (g^{e}_{A})^{2}\right) \right].
\end{equation}
Taking into account that the top quark channel opens at $2m_{t}=350~GeV$, reduces this width $18\%$ below the $\bar{t}t$ threshold, 
but already for $M_{Z^{\prime}}=500~GeV$ this phase space correction is of the order of $2\%$, thus we can safely neglect fermion 
masses.

The cross section for the $Z^{\prime}$ contributions to $l^{+}l^{-}$ production in hadron colliders can be written as \cite{Carena:2004xs}, \cite{Accomando:2010fz}
\begin{equation}
\sigma_{l^{+}l^{-}}=\frac{\pi}{48 s}\left[ c_{u} w_{u}(s, M^{2}_{Z^{\prime}}) + c_{d} w_{d}(s, M^{2}_{Z^{\prime}}) \right],
\end{equation}
where the functions $w_{u,d}(s, M^{2}_{Z^{\prime}})$ depend only on the invariant Mandelstam variable $s$ of the collision and on the 
$Z^{\prime}$ mass and are the same for any model containing neutral gauge bosons with generation-independent couplings to quarks. 
Explicit expressions for these functions in terms of the parton distribution functions of the colliding hadrons can be found in 
\cite{Carena:2004xs}. The coefficients $c_{u,d}$ depend on the $Z^\prime$ couplings to fermions as
\begin{align}
c_{u}&= \frac{g^{\prime 2}}{2} \left[ (g^{u}_{V})^{2} + (g^{u}_{A})^{2}    \right] BR(Z^{\prime} \to l^{+}l^{-} ), 
\label{cu}\\
c_{d}&= \frac{g^{\prime 2}}{2} \left[ (g^{d}_{V})^{2} + (g^{d}_{A})^{2}    \right] BR(Z^{\prime} \to l^{+}l^{-} ).
\label{cd}
\end{align}
Using the form given in \cite{Carena:2004xs} for the functions $w_{u,d}(s, M^{2}_{Z^{\prime}})$, experimental data on $l^{+}l^{-}$ production 
in hadron colliders can be used to impose upper bounds on the intermediate $Z^{\prime}$ contributions which
translates into exclusion curves in the $c_{u}-c_{d}$ plane for given values of $M_{Z^{\prime}}$. In our formalism the $Z^{\prime}$ couplings to
standard model fermions are fixed by known data and the $Z^{\prime}$ mass, thus we are able to calculate the predicted values of $c_{u,d}$ as functions of $M_{Z^{\prime}}$ and to compare these results with the exclusion curves extracted from the data.

\subsection{$Z^{\prime}$ contributions to $l^{+}l^{-}$ production for theories with kinetic mixing and custodial symmetry}

\begin{figure}[h]
\includegraphics[scale=0.4]{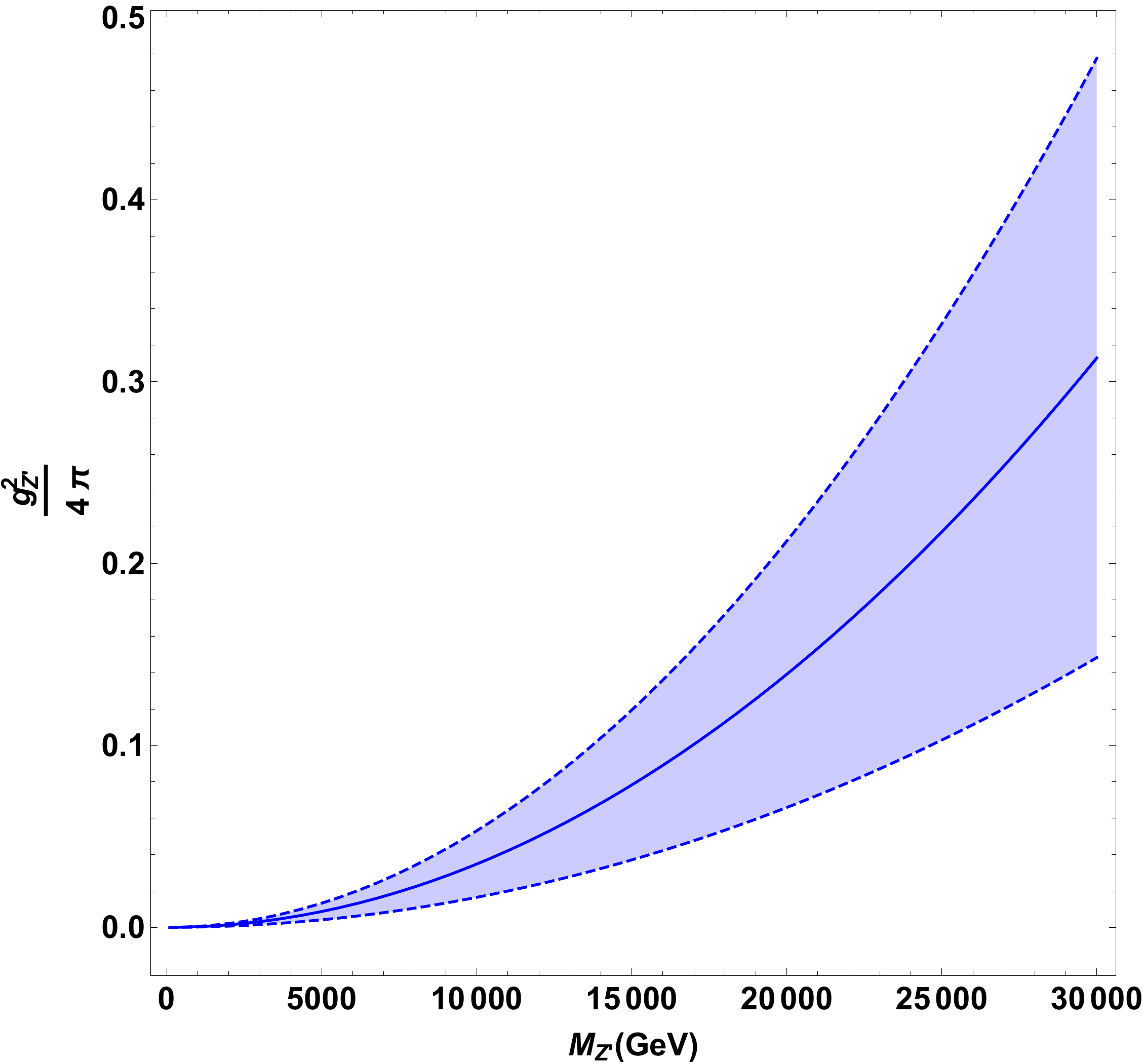}
\caption{Fine structure constant for the coupling $g_{Z^{\prime}}$ induced by kinetic mixing as a function of the $Z^{\prime}$ mass. 
The solid line corresponds to the predictions using the central value of $\rho_0$ and the shadow band to the 
$1\sigma$ region for $\rho_{0}$ in Eq.(\ref{rho0exp}). }
\label{gpplot}
\end{figure}

In the case that SM fermions do not carry the $U(1)_{d}$ charges, the covariant derivative in Eq.(\ref{covder2}) yields the following 
$Z^{\prime}\bar{f}f$ interactions
\begin{equation}
{\cal L}_{Z^{\prime}\bar{f}f}=g_{Z^{\prime}} \sum_{f} \bar{f}\gamma^{\mu}
\left[ T^{3}_{f_{L}} -2 (1-\sigma_{0}\hat{c}^{2}_{Z}) Q - T^{3}_{f_{L}} \gamma^{5} \right]f\, Z^{\prime}_{\mu} ,
\label{lagzprime}
\end{equation}
where
\begin{equation}
g_{Z^{\prime}} = \frac{e}{2\hat{s}_{Z} \hat{c}_{Z}}  \sqrt{\frac{(\rho_0-1)(\rho_{0}\hat{c}^{2}_{Z} - \hat{s}^{2}_{Z}) }
{\hat{c}^{2}_{Z}\sigma_{0}(\rho_{0}-\sigma_{0})}}.
\label{gprime}
\end{equation}
Notice that this coupling scales as $\sqrt{(\rho_{0}-1)/\sigma_{0}}$, thus being small for $Z^{\prime}$ masses close to the electroweak scale, 
but the $\sqrt{1/\sigma_{0}}$ factor enhance it for $M_{Z^{\prime}} >> M_{W}$ and eventually becomes
large for large values of the $Z^{\prime}$ mass. In Fig. \ref{gpplot} we show the associated fine structure constant 
$g^{2}_{Z^{\prime}} /4\pi$ for the window of values at the $1\sigma$ level for $\rho_{0}$ in Eq. (\ref{rho0exp}), which shows that for  
$M_{Z^{\prime}}\sim 30 ~TeV$ we enter in a non-perturbative regime. The couplings of the $Z^{\prime}$ to fermions are 
generation independent but non-universal, depending of the $T^{3}_{L}$ and $Q$ quantum numbers of the specific fermion.

Comparing the interacting Lagrangian in Eq. (\ref{lagzprime}) with the general Lagrangian in Eq. (\ref{lagzpgen}), we obtain 
 $g^{\prime}=g_{Z^{\prime}}$ which according to Eq. (\ref{gprime}) depends only on known data and $M_{Z^{\prime}}$. We also identify
the following vector and axial factors
\begin {align}
g^{f}_{V}&=T^{3}_{f}-2(1-\sigma_{0}\hat{c}^{2}_{Z})Q_{f},\qquad g^{f}_{A}=T^{3}_{f}.
\label{vacoup}
\end{align}
These couplings yield the following total $Z^{\prime}$ decay width into SM fermions
\begin{equation}
\Gamma^{f}_{Z^{\prime}}  =  \frac{\alpha M_{Z^{\prime}}}{4 \hat{s}^{2}_{Z} \hat{c}^{2}_{Z}}
\frac{(\rho_0-1)(\rho_{0}\hat{c}^{2}_{Z} - \hat{s}^{2}_{Z}) }{\hat{c}^{2}_{Z} \sigma_{0}(\rho_{0}-\sigma_{0})}
\left[  1-2(1-\sigma_{0}\hat{c}^{2}_{Z}) + \frac{8}{3} (1-\sigma_{0}\hat{c}^{2}_{Z})^{2}\right].
\label{Gammaf}
\end{equation}

\begin{figure}[h]
\includegraphics[scale=0.5]{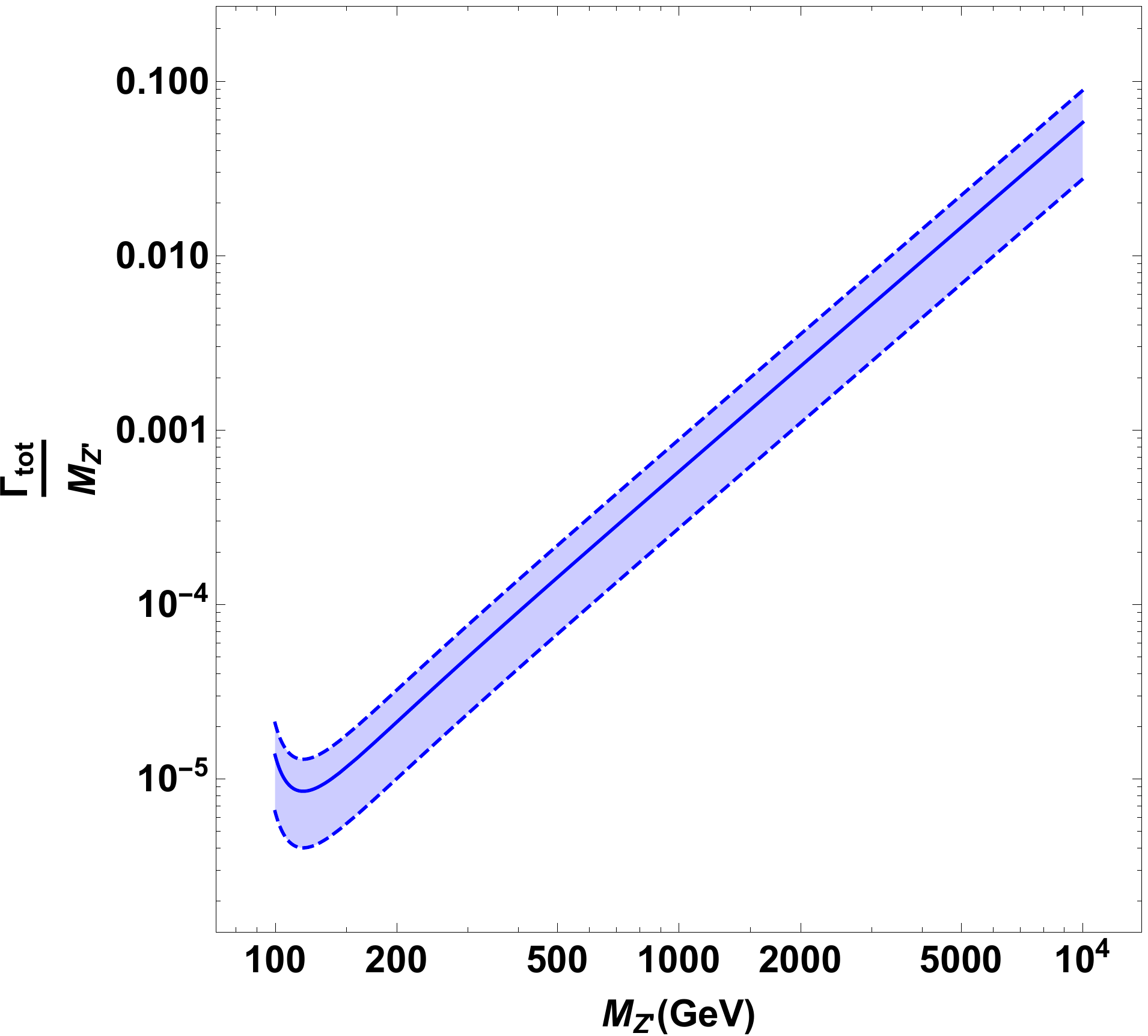}
\caption{Width to mass ratio for the $Z^{\prime}$ boson  as a function of the $Z^{\prime}$ mass. The solid line corresponds 
to the predictions using the central value of $\rho_0$ and the shadow band to the $1\sigma$ region for $\rho_{0}-1$ in Eq.(\ref{rho0m1}). }
\label{nwa}
\end{figure}

The general formalism for the calculation of the $Z^{\prime}$ production in hadron colliders uses the narrow width 
approximation \cite{Carena:2004xs}, \cite{Accomando:2010fz} and we must ensure that 
we are in this regime for the energies at which data is obtained. In a first approximation we will assume that the total width of the 
$Z^{\prime}$ is given by its decays to SM fermions  and will study below possible modifications to this picture. Under this assumption,
$\Gamma^{f}_{Z^{\prime}}$ in Eq.(\ref{Gammaf}) is the total width. In Fig. \ref{nwa} we plot the ratio 
$\Gamma^{f}_{Z^{\prime}}/M_{Z^{\prime}}$ as a function of $M_{Z^{\prime}}$ and we can see that the narrow width approximation is 
well satisfied up to masses of the order of $10~TeV$. At the $M_{Z^{\prime}}=6~TeV$ this ratio is at the $3\%$ level reaching values 
of the order of $10\%$ for $M_{Z^{\prime}}=10.6~TeV$, thus we can use safely the calculations based on the narrow width 
approximation up to this energy. 
The branching ratio for the $l^{+}l^{-}$ channel is obtained as
\begin{equation}
BR(Z^{\prime} \to l^{+}l^{-} ) =  \frac{1}{8} \frac{1-4(1-\sigma_{0}\hat{c}^{2}_{Z}) + 8 (1-\sigma_{0}\hat{c}^{2}_{Z})^{2}}
{  3-6(1-\sigma_{0}\hat{c}^{2}_{Z}) + 8 (1-\sigma_{0}\hat{c}^{2}_{Z})^{2}}.
\label{BR}
\end{equation}
Finally, a calculation of the coefficients $c_{u,d}$ in Eqs.(\ref{cu},\ref{cd}) with $g{\prime}=g_{Z^{\prime}}$ and the vector and axial factors in 
Eq. (\ref{vacoup}) yields
\begin{align}
c_{u}=& \frac{\pi\alpha}{36 \hat{s}^{2}_{Z}\hat{c}^{2}_{Z}} \frac{(\rho_0-1)(\rho_{0}\hat{c}^{2}_{Z} - \hat{s}^{2}_{Z}) }
{\hat{c}^{2}_{Z} \sigma_{0}(\rho_{0} -\sigma_{0})}
 \left[ 9 - 24 (1-\sigma_{0}\hat{c}^{2}_{Z}) + 32 (1-\sigma_{0}\hat{c}^{2}_{Z})^{2}\right] BR(Z^{\prime} \to l^{+}l^{-} ), 
  \label{cukm} \\
c_{d}= &  \frac{\pi\alpha}{36 \hat{s}^{2}_{Z} \hat{c}^{2}_{Z}} \frac{(\rho_0-1)(\rho_{0}\hat{c}^{2}_{Z} - \hat{s}^{2}_{Z}) }
{\hat{c}^{2}_{Z}\sigma_{0}(\rho_{0} -\sigma_{0})}
 \left[ 9 - 12 (1-\sigma_{0}\hat{c}^{2}_{Z}) + 8 (1-\sigma_{0}\hat{c}^{2}_{Z})^{2}\right] BR(Z^{\prime} \to l^{+}l^{-} ).
  \label{cdkm}
\end{align}
Notice that these factors satisfy the linear relation
\begin{equation}
c_{u}=  \frac{ 9 - 24 (1-\sigma_{0}\hat{c}^{2}_{Z}) + 32 (1-\sigma_{0}\hat{c}^{2}_{Z})^{2} }{ 9 - 12 (1-\sigma_{0}\hat{c}^{2}_{Z}) 
+ 8 (1-\sigma_{0}\hat{c}^{2}_{Z})^{2}}  c_{d}
\end{equation}
independently of the value of $ BR(Z^{\prime} \to l^{+}l^{-} )$.  In general the proportionality coefficient depends on $M^{2}_{Z^{\prime}}$, but for
 large $Z^{\prime}$ mass, it reaches an asymptotic constant value to yield
 \begin{equation}
c_{u}=  \frac{ 17}{ 5}  c_{d}.
\end{equation}
 In this limit, the branching ratio in Eq.(\ref{BR}) reaches a saturation value $BR(l^{+}l^{-})=1/8$ and the values of the $c_{u,d}$
factors grow like $M^{2}_{Z^{\prime}}$ 
\begin{align}
c_{u}\approx&\frac{17}{8} \frac{\pi\alpha}{36 \hat{s}^{2}_{Z} \hat{c}^{2}_{Z}} \frac{(\rho_{0}\hat{c}^{2}_{Z} - \hat{s}^{2}_{Z}) }{\rho_{0}} 
 \frac{(\rho_0-1)}{\hat{c}^{2}_{Z} \sigma_{0}} = 1.67 \times 10^{-6} \frac{M^{2}_{Z^{\prime}}}{M^{2}_{W}} , \\
c_{d}\approx & \frac{5}{8}  \frac{\pi\alpha}{36 \hat{s}^{2}_{Z} \hat{c}^{2}_{Z} } \frac{(\rho_{0}\hat{c}^{2}_{Z} - \hat{s}^{2}_{Z}) }{\rho_{0}} 
 \frac{(\rho_0-1)}{\hat{c}^{2}_{Z} \sigma_{0}}=  0.49 \times 10^{-6} \frac{M^{2}_{Z^{\prime}}}{M^{2}_{W}}.
\end{align}
In practice, large $Z^{\prime}$ mass means $\sigma_{0}<< 1$, which is satisfied for $M_Z{^{\prime}}\gtrsim 500 ~GeV$. 

\begin{figure}[h]
\includegraphics[scale=0.3]{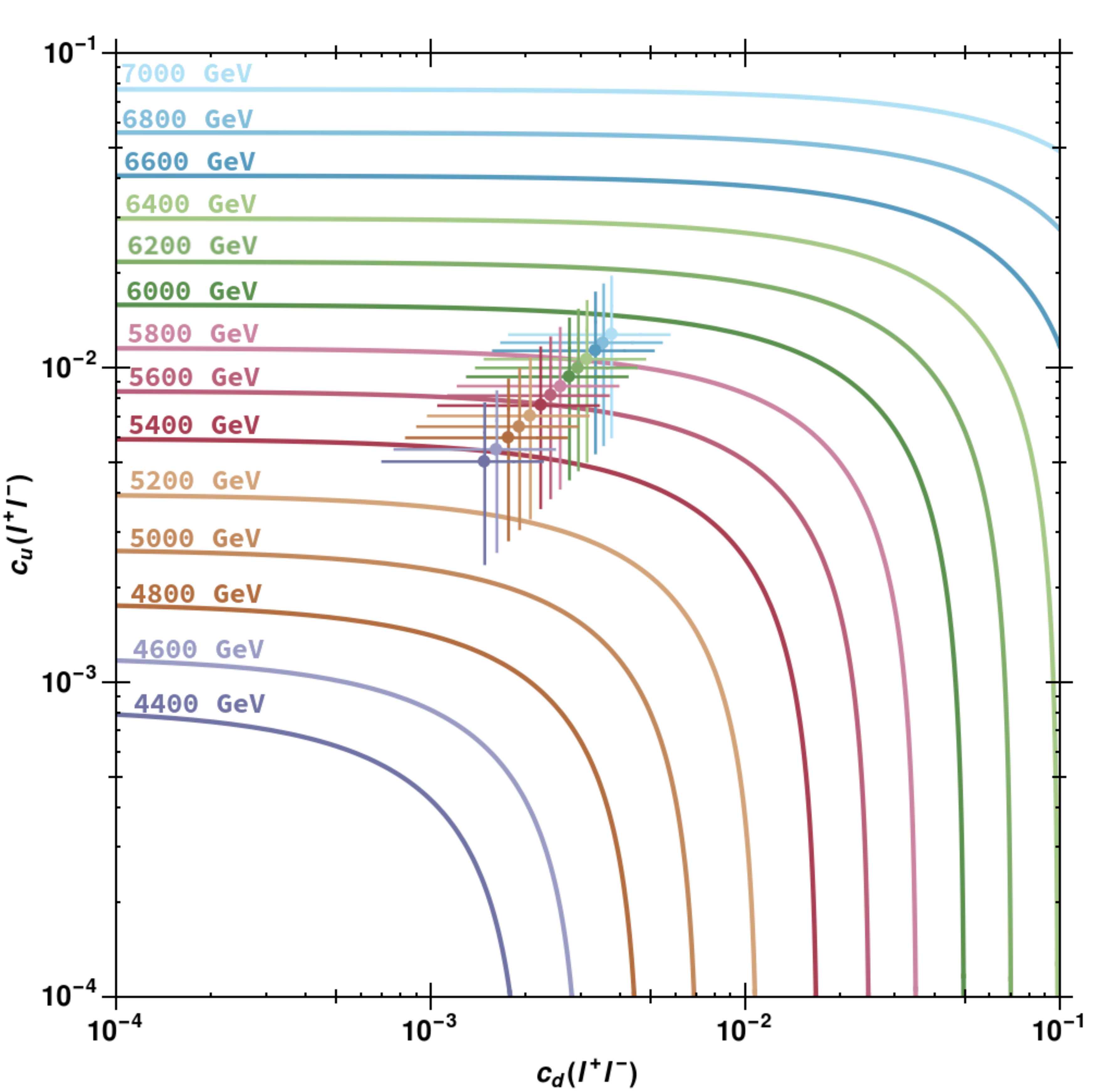}
\caption{Exclusion curves for the $c_{u},c_{d}$ couplings extracted from Ref.  \cite{CMS:2021ctt} and the corresponding predictions in theories 
for physics BSM containing an extra spontaneously broken $U(1)_{d}$ and respecting custodial symmetry. Each dot corresponds to the 
predicted value for a given $Z^{\prime}$ boson mass from Eqs. (\ref{cukm},\ref{cdkm}) and is marked in the same color as the 
corresponding exclusion curve. Uncertainties correspond to the $1\sigma$ region for $\rho_{0}$ in Eq. (\ref{rho0exp}).  }
\label{cucdplot}
\end{figure}

Experimental data on the upper bounds for $Z^{\prime}$ production at the LHC has been translated into exclusion curves  in the 
$c_{d}-c_{u}$ plane for given values of $M_{Z^{\prime}}$. In Fig. \ref{cucdplot} we show the exclusion curves obtained in  
\cite{CMS:2021ctt}  and our results. The dots in this plot are the values of 
$(c_{d}(M_{Z^{\prime}}),c_{u}(M_{Z^{\prime}}))$ from Eqs. (\ref{cukm},\ref{cdkm}) for the values of $M_{Z^{\prime}}$ corresponding 
to the exclusion curves and are computed taking the central value of $\rho_{0}$ in Eq.(\ref{rho0exp}). The dot for a given mass is marked 
in the same color as the corresponding exclusion curve. Uncertainties in this plot correspond to the $1\sigma$ region for $\rho_{0}$ in 
Eq.(\ref{rho0exp}).  For $M_Z{^{\prime}}< 5200~ GeV$ the predicted values for $c_{u}, c_{d}$ including uncertainties are above the 
exclusion curves thus, at $1\sigma$ level, the predicted values of the parameters are inconsistent with data.  

A more precise calculation requires to take into account the uncertainties in $c_{u},c_{d}$ induced by changes in 
$BR(Z^{\prime} \to l^{+}l^{-} )$ due to contributions of other decay channels to the total decay width.  
First, although the mixing with the SM fields generates couplings of the $Z^{\prime}$ to $ZZ$, $W^{+}W^{-}$ 
and $ZH$, these couplings are proportional to single mixing factors $s_{z}$ and the corresponding decay widths are proportional to 
$\rho_{0}-1$, thus small compared to the decay widths to fermions which are enhanced by the $1/\sigma_{0}$ factor and turn out to be 
proportional to $(\rho_{0}-1)M^{2}_{Z^\prime}/M^{2}_{W}$. Second, the decay to non-SM particles in the ultraviolet completing theory 
may be more important since, as we can see from Eqs. (\ref{covder2},\ref{czsecchi}), the corresponding coupling has the same 
enhancement factor as the coupling to fermions and we must have at least a rough estimate of the decay width to these non-SM particles. 
In this concern, the most important physical case today is the possibility that dark matter enter particle physics coupled to SM fields 
via kinetic mixing, in which case, it is natural to expect dark matter particles with masses of the order of the electroweak scale. 

\begin{figure}[ht]
\includegraphics[scale=0.6]{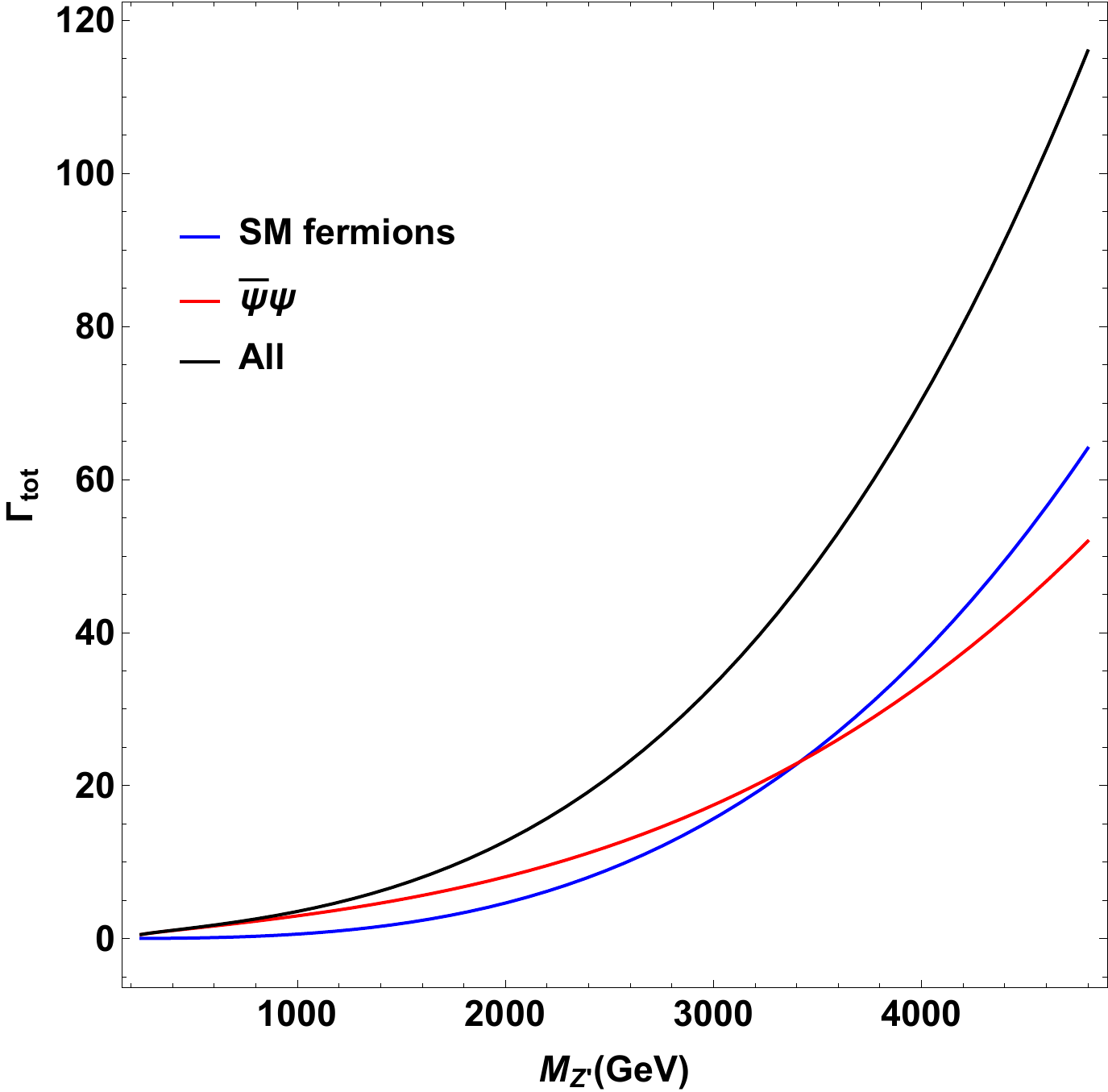}
\caption{Estimate of the decay width of the $Z^{\prime}$ to extra fermions with a mass of the order of the electroweak scale and a 
perturbative coupling.   }
\label{Gfplot}
\end{figure}

In order to have an estimate of these effects we consider the possibility of extra fermion field $\psi$ as (dark) matter field with $U(1)_{d}$ 
charge $Q_{v}^{\psi}=2$, such that its coupling to the $V^{\mu}$ field in Eq. (\ref{Lag}) is $g_{v}$.  In this case, the $Z^{\prime}\bar{\psi}\psi$ 
interaction arising from the covariant derivative in Eq. (\ref{covder2}) is
\begin{equation}
{\cal L}_{Z^{\prime}\bar{\psi}\psi}=- g_{v}\hat{c}_{\zeta} \sec\hat\chi \bar{\psi}\gamma^{\mu}\psi Z^{\prime}_{\mu},
\end{equation}
which yields the following decay width
\begin{align}
\Gamma( Z^{\prime}\to\bar{\psi} \psi ) &=  \frac{\alpha_{v}M_{Z^{\prime}}}{3 \hat{s}^{2}_{Z} \hat{c}^{2}_{Z}}
\left[ \frac{\rho_{0}- \hat{s}^{2}_{Z}-\rho_{0}\sigma_{0} \hat{c}^{2}_{Z} }{\rho_{0}-\sigma_{0}}
\left( \frac{(\rho_{0}-1)}{\sigma_{0}} \frac{(\rho_{0}(1-\sigma_{0})\hat{c}^{2}_{Z}- \hat{s}^{2}_{Z})}{\rho_{0}\hat{c}^{2}_{Z}}  
+ \rho_{0} \hat{s}^{2}_{Z} \right)  
\right] \nonumber \\
&\times \left(1+\frac{2 M^{2}_{\psi}}{M^{2}_{Z^{\prime}}}\right)\sqrt{ 1-\frac{4 M^{2}_{\psi}}{M^{2}_{Z^{\prime}}}},
\end{align}
where $\alpha_{v}=g^{2}_{v}/4\pi$ is the $U(1)_{d}$ fine structure constant. This width depends on the unknown $\alpha_{v}$, $M_{\psi}$ 
and $M_{Z^{\prime}}$ and it is not possible to fix it with certainty. However, a reasonable estimate of its size can be obtained 
considering that the $U(1)_{d}$ interaction is perturbative and, at least for the dark matter context we can consider 
$M_{\psi}$ of the order of the electroweak scale. For larger masses, phase space reduces for a given value of the $Z^{\prime}$ mass 
and we expect a smaller decay width.   
In Fig. \ref{Gfplot} we show the decay width for $Z^{\prime}\to\bar{\psi} \psi$ taking $\alpha_{v}=\alpha$ and $M=100~ GeV$, as a 
function of $M_{Z^{\prime}}$ together with the result of the decay width to fermions in Eq.(\ref{Gammaf}). It is clear from this plot that 
around $M_{Z^{\prime}}=5.2~TeV$ this contribution is of the same order as the decay width to SM fermions, thus we expect the values 
of $(c_{d},c_{u})$ to decrease roughly by a factor of $1/2$. Taking this contribution into account we find that the narrow width 
approximation is valid up to energies of $10~TeV$, where the width to mass ratio reaches the $10\%$ level. 
Considering the $1/2$ factor correction to our previous calculation and the uncertainties in the value of $\rho_{0}$ in Eq.(\ref{rho0exp}), 
from the exclusion curves given in  \cite{CMS:2021ctt} we refine the lower limit to $M_{Z^\prime}\gtrsim 5.0 ~TeV$. 

\section{ Conclusions}
In this work we study the implications of kinetic mixing for the $Z$ and $Z^{\prime}$ interactions in a class of models for physics 
beyond the SM motivated by the dark matter problem, whose gauge group contains a spontaneously broken $U(1)_{d}$ factor 
group and whose Higgs sector respects custodial symmetry. 
For these theories, the kinetic mixing generates a coupling of the extra neutral physical gauge boson $Z^{\prime}$ to SM fermions. 
In a hidden dark matter scenario, custodial symmetry allows to write the kinetic mixing parameters in terms of the measured 
values of $M_{Z}$, the electromagnetic constant $\alpha$, the Fermi constant $G_{F}$ and the mass of the $Z^\prime$ 
\cite{Napsuciale:2020kai}, in such a way that the $Z$ and $Z^{\prime}$ couplings to SM fermions can be written in terms 
of these parameters. 

On the other hand, the fit to electroweak precision data requires the calculation 
of observables at the loop level and encode possible effects of physics beyond the standard model in the deviation of the 
$\rho_{0}=M^{2}_{W}/\hat{c}^{2}_{Z}M^{2}_{Z}\hat{\rho}$ from the unit value. The value  of $\rho_{0}$ is 
fixed to $\rho_{0}=1.00038\pm 0.00020$ from the global fit to electroweak precision data \cite{Zyla:2020zbs} thus 
$\rho_{0}>1$ at $1.9~\sigma$ level ($94\%$ confidence level) and although not conclusive, present data 
points to the existence of  new physics contributions to the $\rho_{0}$ parameter.

Custodial symmetry protects the equal mass relation for $\bf{W}_{\mu}$ triplet from radiative corrections. This and the unbroken 
nature of the electromagnetic $U(1)_{em}$ group generated by $Q=T_{3}+Y/2$ allows to relate the $Z$ add $W^{\pm}$ masses 
at the loop level in the standard model as $M^{2}_{W}=M^{2}_{Z}\hat{c}^{2}_{Z}\hat{\rho}$, where $\hat{\rho}$ are small radiative 
corrections arising from custodial symmetry violating Yukawa interactions which enter the mass relations due to the mixing 
(characterized by the weak mixing angle) of the $W^{3}_{\mu}$ with the $U(1)_{Y}$ gauge boson $B_{\mu}$ to produce the 
standard model $Z_{\mu}$ and the photon field $A_{\mu}$.
 
In an extension of the standard model with a spontaneously broken $U(1)_{d}$ factor dark gauge group and kinetic mixing, 
if the Higgs sector respects custodial symmetry, then small couplings between the SM and dark sectors are generated, new 
mixing terms appear and mixing pattern in the neutral gauge sector is more involved. However, in the hidden dark matter 
scenario, custodial symmetry still relates the $W^{3}_{\mu}$ mass term to the SM $Z_{\mu}$ (denoted $\tilde{Z}_{\mu}$ here) 
mass term, although now $\tilde{Z}_{\mu}$ is not the physical field. We argue that at the $\mu=M_{Z}$ scale, radiative corrections 
are dominated by SM particles and use this relation to calculate at the loop level the $Z$ and $Z^{\prime}$ couplings and to write 
them in terms of the measured values of $M_{Z}$, $\alpha$, $G_{F}$ and the mass of the $Z^\prime$. Even if radiative corrections 
involving contributions of the $Z^{\prime}$ are small, there is a tree-level mixing $\tilde{Z} - \bar{V}$ mixing which modifies the 
naive custodial symmetry relation such that in the extended theory $\rho_{0}\neq 1$. We calculate $\rho_{0}$ in the extended 
theory and use it to rewrite the $Z$ and $Z^{\prime}$ couplings in terms of $\rho_{0}$ and the analogous ratio 
$\sigma_{0}=M^{2}_{W}/\hat{c}^{2}_{Z}M^{2}_{Z^{\prime}}\hat{\rho}$. We use these couplings to calculate some observables 
where physics beyond the SM may produce some effects at the electroweak scale.

For the $Z$ boson we calculate the oblique parameters $S$ and $T$, which encode the low energy effects of physics beyond the SM 
for a wide class of theories, including the ones considered here.  We find that for $M_{Z^{\prime}}> 200 ~GeV$ the oblique 
parameters are not sensitive to the $Z^{\prime}$ mass and a comparison with the values of $S$ and $T$ extracted from the global 
fit to electroweak precision data \cite{Zyla:2020zbs} at $1\sigma$ level, yields results in agreement with the fit for 
$M_{Z^{\prime}}>M_{Z}$. This lower bound is consistent with results found in \cite{Napsuciale:2020kai} based on the physical range 
of values for the mixing angles. 

As to the $Z^{\prime}$ physics we study the intermediate $Z^{\prime}$ contributions to the production of a charged lepton pair at the LHC. 
The corresponding cross section can written in terms of two parameters, $c_{u}$ and $c_{d}$, 
carrying all the information of the $Z^{\prime}$ couplings to SM fermions. There are two factors in $c_{u}$ and $c_{d}$, the first of which 
contains only the couplings to SM fermions. The second factor is the branching ratio $BR(Z^{\prime}\to l^{+}l^{-})$ which contains 
information on SM and beyond the SM physics. In a first approximation, we calculate the $c_{u}$ and $c_{d}$ parameters in our 
formalism considering only the $Z^{\prime}$ couplings to SM fermions generated by the kinetic mixing. In this case, the corresponding 
cross section depends on the unknown $Z^{\prime}$ mass and known SM parameters. We compare the results of our calculation 
using the range of values of $\rho_{0}$ extracted from the global fit to the EWPD at the $1\sigma$ level 
with the exclusion curves in the $c_{u}-c_{d}$ plane for $Z^{\prime}$ masses in the range $3.8-7.0~TeV$ obtained by the CMS Collaboration  
\cite{Sirunyan:2018exx}, \cite{CMS:2021ctt} . This comparison shows that consistency of our calculation with the CMS data requires 
$M_{Z^{\prime}}\geq 5.2 ~TeV$. This result is modified when we consider the coupling of the $Z^{\prime}$ to kinematically allowed non-SM 
particles which enter $c_{u}$ and $c_{d}$ through the branching ratio $BR(Z^{\prime}\to l^{+}l^{-})$. We estimate these contributions in 
the well motivated case of  dark matter entering particle physics as the matter fields of the $U(1)_{d}$ gauge symmetry with perturbative 
couplings at the electroweak scale, finding that our results for  $c_{u}$ and $c_{d}$ get modified roughly by a factor of $1/2$. Taking into 
account this correction we obtain that consistency with the CMS data requires $M_{Z^{\prime}}\gtrsim 5.0~ TeV$. 
These lower limits are obtained considering the $1\sigma$ region ($68\%$ confidence level) for the values of $\rho_{0}$ extracted from the fit 
to EWPD.  Although present data points to the existence of  new physics 
contributions to the $\rho_{0}$ parameter, definitive conclusions must await for more precise electroweak data which will eventually allow 
to lower the uncertainty in this parameter. 
 
\section{Acknowledgments}
One of us (H.H.A.) acknowledges CONACyT-M\'{e}xico for a scholarship to pursue her Ph. D. .

\bibliographystyle{JHEP}
\bibliography{Zprime}

\providecommand{\href}[2]{#2}\begingroup\raggedright\begin{thebibliography}{10}

\bibitem{Hewett:1988xc}
J.~L. Hewett and T.~G. Rizzo, \emph{{Low-Energy Phenomenology of Superstring
  Inspired E(6) Models}},
  \href{https://doi.org/10.1016/0370-1573(89)90071-9}{\emph{Phys. Rept.}
  {\bfseries 183} (1989) 193}.

\bibitem{Langacker:2008yv}
P.~Langacker, \emph{{The Physics of Heavy $Z^\prime$ Gauge Bosons}},
  \href{https://doi.org/10.1103/RevModPhys.81.1199}{\emph{Rev. Mod. Phys.}
  {\bfseries 81} (2009) 1199}
  [\href{https://arxiv.org/abs/0801.1345}{{\ttfamily 0801.1345}}].

\bibitem{Holdom:1985ag}
B.~Holdom, \emph{{Two U(1)'s and Epsilon Charge Shifts}},
  \href{https://doi.org/10.1016/0370-2693(86)91377-8}{\emph{Phys. Lett. B}
  {\bfseries 166} (1986) 196}.

\bibitem{Dienes:1996zr}
K.~R. Dienes, C.~F. Kolda and J.~March-Russell, \emph{{Kinetic mixing and the
  supersymmetric gauge hierarchy}},
  \href{https://doi.org/10.1016/S0550-3213(97)00173-9}{\emph{Nucl. Phys. B}
  {\bfseries 492} (1997) 104}
  [\href{https://arxiv.org/abs/hep-ph/9610479}{{\ttfamily hep-ph/9610479}}].

\bibitem{Babu:1997st}
K.~Babu, C.~F. Kolda and J.~March-Russell, \emph{{Implications of generalized Z
  - Z-prime mixing}},
  \href{https://doi.org/10.1103/PhysRevD.57.6788}{\emph{Phys. Rev. D}
  {\bfseries 57} (1998) 6788}
  [\href{https://arxiv.org/abs/hep-ph/9710441}{{\ttfamily hep-ph/9710441}}].

\bibitem{Babu:1996vt}
K.~Babu, C.~F. Kolda and J.~March-Russell, \emph{{Leptophobic U(1) $s$ and the
  R($b$) - R($c$) crisis}},
  \href{https://doi.org/10.1103/PhysRevD.54.4635}{\emph{Phys. Rev. D}
  {\bfseries 54} (1996) 4635}
  [\href{https://arxiv.org/abs/hep-ph/9603212}{{\ttfamily hep-ph/9603212}}].

\bibitem{Baumgart:2009tn}
M.~Baumgart, C.~Cheung, J.~T. Ruderman, L.-T. Wang and I.~Yavin,
  \emph{{Non-Abelian Dark Sectors and Their Collider Signatures}},
  \href{https://doi.org/10.1088/1126-6708/2009/04/014}{\emph{JHEP} {\bfseries
  04} (2009) 014} [\href{https://arxiv.org/abs/0901.0283}{{\ttfamily
  0901.0283}}].

\bibitem{Cheung:2009qd}
C.~Cheung, J.~T. Ruderman, L.-T. Wang and I.~Yavin, \emph{{Kinetic Mixing as
  the Origin of Light Dark Scales}},
  \href{https://doi.org/10.1103/PhysRevD.80.035008}{\emph{Phys. Rev. D}
  {\bfseries 80} (2009) 035008}
  [\href{https://arxiv.org/abs/0902.3246}{{\ttfamily 0902.3246}}].

\bibitem{Ibarra:2009bm}
A.~Ibarra, A.~Ringwald, D.~Tran and C.~Weniger, \emph{{Cosmic Rays from
  Leptophilic Dark Matter Decay via Kinetic Mixing}},
  \href{https://doi.org/10.1088/1475-7516/2009/08/017}{\emph{JCAP} {\bfseries
  08} (2009) 017} [\href{https://arxiv.org/abs/0903.3625}{{\ttfamily
  0903.3625}}].

\bibitem{Hook:2010tw}
A.~Hook, E.~Izaguirre and J.~G. Wacker, \emph{{Model Independent Bounds on
  Kinetic Mixing}}, \href{https://doi.org/10.1155/2011/859762}{\emph{Adv. High
  Energy Phys.} {\bfseries 2011} (2011) 859762}
  [\href{https://arxiv.org/abs/1006.0973}{{\ttfamily 1006.0973}}].

\bibitem{Chun:2010ve}
E.~J. Chun, J.-C. Park and S.~Scopel, \emph{{Dark matter and a new gauge boson
  through kinetic mixing}},
  \href{https://doi.org/10.1007/JHEP02(2011)100}{\emph{JHEP} {\bfseries 02}
  (2011) 100} [\href{https://arxiv.org/abs/1011.3300}{{\ttfamily 1011.3300}}].

\bibitem{Mambrini:2010dq}
Y.~Mambrini, \emph{{The Kinetic dark-mixing in the light of CoGENT and
  XENON100}}, \href{https://doi.org/10.1088/1475-7516/2010/09/022}{\emph{JCAP}
  {\bfseries 09} (2010) 022} [\href{https://arxiv.org/abs/1006.3318}{{\ttfamily
  1006.3318}}].

\bibitem{Mambrini:2011dw}
Y.~Mambrini, \emph{{The ZZ' kinetic mixing in the light of the recent direct
  and indirect dark matter searches}},
  \href{https://doi.org/10.1088/1475-7516/2011/07/009}{\emph{JCAP} {\bfseries
  07} (2011) 009} [\href{https://arxiv.org/abs/1104.4799}{{\ttfamily
  1104.4799}}].

\bibitem{Brahmachari:2014aya}
B.~Brahmachari and A.~Raychaudhuri, \emph{{Kinetic mixing and symmetry breaking
  dependent interactions of the dark photon}},
  \href{https://doi.org/10.1016/j.nuclphysb.2014.08.015}{\emph{Nucl. Phys. B}
  {\bfseries 887} (2014) 441}
  [\href{https://arxiv.org/abs/1409.2082}{{\ttfamily 1409.2082}}].

\bibitem{Arguelles:2016ney}
C.~A. Arguelles, X.-G. He, G.~Ovanesyan, T.~Peng and M.~J. Ramsey-Musolf,
  \emph{{Dark Gauge Bosons: LHC Signatures of Non-Abelian Kinetic Mixing}},
  \href{https://doi.org/10.1016/j.physletb.2017.04.037}{\emph{Phys. Lett. B}
  {\bfseries 770} (2017) 101}
  [\href{https://arxiv.org/abs/1604.00044}{{\ttfamily 1604.00044}}].

\bibitem{Belanger:2017vpq}
G.~Belanger, J.~Da~Silva and H.~M. Tran, \emph{{Dark matter in U(1) extensions
  of the MSSM with gauge kinetic mixing}},
  \href{https://doi.org/10.1103/PhysRevD.95.115017}{\emph{Phys. Rev. D}
  {\bfseries 95} (2017) 115017}
  [\href{https://arxiv.org/abs/1703.03275}{{\ttfamily 1703.03275}}].

\bibitem{Arcadi:2018tly}
G.~Arcadi, T.~Hugle and F.~S. Queiroz, \emph{{The Dark $L_\mu - L_\tau$ Rises
  via Kinetic Mixing}},
  \href{https://doi.org/10.1016/j.physletb.2018.07.028}{\emph{Phys. Lett. B}
  {\bfseries 784} (2018) 151}
  [\href{https://arxiv.org/abs/1803.05723}{{\ttfamily 1803.05723}}].

\bibitem{Foot:2012ai}
R.~Foot, \emph{{Implications of mirror dark matter kinetic mixing for CMB
  anisotropies}},
  \href{https://doi.org/10.1016/j.physletb.2012.12.001}{\emph{Phys. Lett. B}
  {\bfseries 718} (2013) 745}
  [\href{https://arxiv.org/abs/1208.6022}{{\ttfamily 1208.6022}}].

\bibitem{Kamada:2018kmi}
A.~Kamada, M.~Yamada and T.~T. Yanagida, \emph{{Self-interacting dark matter
  with a vector mediator: kinetic mixing with the $
  \mathrm{U}{(1)}_{{\left(B-L\right)}_3} $ gauge boson}},
  \href{https://doi.org/10.1007/JHEP03(2019)021}{\emph{JHEP} {\bfseries 03}
  (2019) 021} [\href{https://arxiv.org/abs/1811.02567}{{\ttfamily
  1811.02567}}].

\bibitem{Rizzo:2018ntg}
T.~G. Rizzo, \emph{{Kinetic mixing, dark photons and an extra dimension. Part
  I}}, \href{https://doi.org/10.1007/JHEP07(2018)118}{\emph{JHEP} {\bfseries
  07} (2018) 118} [\href{https://arxiv.org/abs/1801.08525}{{\ttfamily
  1801.08525}}].

\bibitem{Rizzo:2018joy}
T.~G. Rizzo, \emph{{Kinetic mixing, dark photons and extra dimensions. Part II:
  fermionic dark matter}},
  \href{https://doi.org/10.1007/JHEP10(2018)069}{\emph{JHEP} {\bfseries 10}
  (2018) 069} [\href{https://arxiv.org/abs/1805.08150}{{\ttfamily
  1805.08150}}].

\bibitem{Rizzo:2018vlb}
T.~G. Rizzo, \emph{{Kinetic Mixing and Portal Matter Phenomenology}},
  \href{https://doi.org/10.1103/PhysRevD.99.115024}{\emph{Phys. Rev. D}
  {\bfseries 99} (2019) 115024}
  [\href{https://arxiv.org/abs/1810.07531}{{\ttfamily 1810.07531}}].

\bibitem{Banerjee:2019asa}
A.~Banerjee, G.~Bhattacharyya, D.~Chowdhury and Y.~Mambrini, \emph{{Dark matter
  seeping through dynamic gauge kinetic mixing}},
  \href{https://doi.org/10.1088/1475-7516/2019/12/009}{\emph{JCAP} {\bfseries
  12} (2019) 009} [\href{https://arxiv.org/abs/1905.11407}{{\ttfamily
  1905.11407}}].

\bibitem{Rueter:2019wdf}
T.~D. Rueter and T.~G. Rizzo, \emph{{Towards A UV-Model of Kinetic Mixing and
  Portal Matter}},
  \href{https://doi.org/10.1103/PhysRevD.101.015014}{\emph{Phys. Rev. D}
  {\bfseries 101} (2020) 015014}
  [\href{https://arxiv.org/abs/1909.09160}{{\ttfamily 1909.09160}}].

\bibitem{Akerib:2019diq}
{\scshape LUX} collaboration, \emph{{First direct detection constraint on
  mirror dark matter kinetic mixing using LUX 2013 data}},
  \href{https://doi.org/10.1103/PhysRevD.101.012003}{\emph{Phys. Rev. D}
  {\bfseries 101} (2020) 012003}
  [\href{https://arxiv.org/abs/1908.03479}{{\ttfamily 1908.03479}}].

\bibitem{Lao:2020inc}
J.~Lao, C.~Cai, Z.-H. Yu, Y.-P. Zeng and H.-H. Zhang, \emph{{Fermionic and
  scalar dark matter with hidden $\mathrm{U}(1)$ gauge interaction and kinetic
  mixing}}, \href{https://doi.org/10.1103/PhysRevD.101.095031}{\emph{Phys. Rev.
  D} {\bfseries 101} (2020) 095031}
  [\href{https://arxiv.org/abs/2003.02516}{{\ttfamily 2003.02516}}].

\bibitem{Gehrlein:2019iwl}
J.~Gehrlein and M.~Pierre, \emph{{A testable hidden-sector model for Dark
  Matter and neutrino masses}},
  \href{https://doi.org/10.1007/JHEP02(2020)068}{\emph{JHEP} {\bfseries 02}
  (2020) 068} [\href{https://arxiv.org/abs/1912.06661}{{\ttfamily
  1912.06661}}].

\bibitem{Kribs:2020vyk}
G.~D. Kribs, D.~McKeen and N.~Raj, \emph{{Breaking up the Proton: An Affair
  with Dark Forces}},
  \href{https://doi.org/10.1103/PhysRevLett.126.011801}{\emph{Phys. Rev. Lett.}
  {\bfseries 126} (2021) 011801}
  [\href{https://arxiv.org/abs/2007.15655}{{\ttfamily 2007.15655}}].

\bibitem{Binh:2020xtf}
D.~Binh, V.~Binh and H.~Long, \emph{{Bounds on Dipole Moments of hidden Dark
  Matter through kinetic mixing}},
  \href{https://arxiv.org/abs/2006.09020}{{\ttfamily 2006.09020}}.

\bibitem{Barnes:2020vsc}
P.~Barnes, Z.~Johnson, A.~Pierce and B.~Shakya, \emph{{Simple Hidden Sector
  Dark Matter}}, \href{https://doi.org/10.1103/PhysRevD.102.075019}{\emph{Phys.
  Rev. D} {\bfseries 102} (2020) 075019}
  [\href{https://arxiv.org/abs/2003.13744}{{\ttfamily 2003.13744}}].

\bibitem{Weinberg:1975gm}
S.~Weinberg, \emph{{Implications of Dynamical Symmetry Breaking}},
  \href{https://doi.org/10.1103/PhysRevD.19.1277}{\emph{Phys. Rev. D}
  {\bfseries 13} (1976) 974}.

\bibitem{Susskind:1978ms}
L.~Susskind, \emph{{Dynamics of Spontaneous Symmetry Breaking in the
  Weinberg-Salam Theory}},
  \href{https://doi.org/10.1103/PhysRevD.20.2619}{\emph{Phys. Rev. D}
  {\bfseries 20} (1979) 2619}.

\bibitem{Sikivie:1980hm}
P.~Sikivie, L.~Susskind, M.~B. Voloshin and V.~I. Zakharov, \emph{{Isospin
  Breaking in Technicolor Models}},
  \href{https://doi.org/10.1016/0550-3213(80)90214-X}{\emph{Nucl. Phys. B}
  {\bfseries 173} (1980) 189}.

\bibitem{Zyla:2020zbs}
{\scshape Particle Data Group} collaboration, \emph{{Review of Particle
  Physics}}, \href{https://doi.org/10.1093/ptep/ptaa104}{\emph{PTEP} {\bfseries
  2020} (2020) 083C01}.

\bibitem{Langacker:1991pg}
P.~Langacker and M.-x. Luo, \emph{{Constraints on additional $Z$ bosons}},
  \href{https://doi.org/10.1103/PhysRevD.45.278}{\emph{Phys. Rev. D} {\bfseries
  45} (1992) 278}.

\bibitem{Napsuciale:2020kai}
M.~Napsuciale, S.~Rodr\'{i}guez and H.~Hern\'{a}ndez-Arellano, \emph{{Kinetic
  mixing, custodial symmetry and a lower bound on the dark $Z^{\prime}$ mass}},
   \href{https://arxiv.org/abs/2008.07051}{{\ttfamily 2008.07051}}.

\bibitem{Lynn:1985fg}
B.~Lynn, M.~E. Peskin and R.~Stuart, \emph{{Radiative Corrections in $SU(2)
  \otimes U(1)$: LEP / SLC}},  7, 1985.

\bibitem{Kennedy:1988sn}
D.~Kennedy and B.~Lynn, \emph{{Electroweak Radiative Corrections with an
  Effective Lagrangian: Four Fermion Processes}},
  \href{https://doi.org/10.1016/0550-3213(89)90483-5}{\emph{Nucl. Phys. B}
  {\bfseries 322} (1989) 1}.

\bibitem{Kuroda:1990wn}
M.~Kuroda, G.~Moultaka and D.~Schildknecht, \emph{{Direct one loop
  renormalization of SU(2)-L x U(1)-Y four fermion processes and running
  coupling constants}},
  \href{https://doi.org/10.1016/0550-3213(91)90252-S}{\emph{Nucl. Phys. B}
  {\bfseries 350} (1991) 25}.

\bibitem{Peskin:1990zt}
M.~E. Peskin and T.~Takeuchi, \emph{{A New constraint on a strongly interacting
  Higgs sector}}, \href{https://doi.org/10.1103/PhysRevLett.65.964}{\emph{Phys.
  Rev. Lett.} {\bfseries 65} (1990) 964}.

\bibitem{Peskin:1991sw}
M.~E. Peskin and T.~Takeuchi, \emph{{Estimation of oblique electroweak
  corrections}}, \href{https://doi.org/10.1103/PhysRevD.46.381}{\emph{Phys.
  Rev. D} {\bfseries 46} (1992) 381}.

\bibitem{Sirunyan:2018exx}
{\scshape CMS} collaboration, \emph{{Search for high-mass resonances in
  dilepton final states in proton-proton collisions at $\sqrt{s}=$ 13 TeV}},
  \href{https://doi.org/10.1007/JHEP06(2018)120}{\emph{JHEP} {\bfseries 06}
  (2018) 120} [\href{https://arxiv.org/abs/1803.06292}{{\ttfamily
  1803.06292}}].

\bibitem{CMS:2021ctt}
{\scshape CMS} collaboration, \emph{{Search for resonant and nonresonant new
  phenomena in high-mass dilepton final states at $ \sqrt{s} $ = 13 TeV}},
  \href{https://doi.org/10.1007/JHEP07(2021)208}{\emph{JHEP} {\bfseries 07}
  (2021) 208} [\href{https://arxiv.org/abs/2103.02708}{{\ttfamily
  2103.02708}}].

\bibitem{Holdom:1990xp}
B.~Holdom, \emph{{Oblique electroweak corrections and an extra gauge boson}},
  \href{https://doi.org/10.1016/0370-2693(91)90836-F}{\emph{Phys. Lett. B}
  {\bfseries 259} (1991) 329}.

\bibitem{Golden:1990ig}
M.~Golden and L.~Randall, \emph{{Radiative Corrections to Electroweak
  Parameters in Technicolor Theories}},
  \href{https://doi.org/10.1016/0550-3213(91)90614-4}{\emph{Nucl. Phys. B}
  {\bfseries 361} (1991) 3}.

\bibitem{Altarelli:1990zd}
G.~Altarelli and R.~Barbieri, \emph{{Vacuum polarization effects of new physics
  on electroweak processes}},
  \href{https://doi.org/10.1016/0370-2693(91)91378-9}{\emph{Phys. Lett. B}
  {\bfseries 253} (1991) 161}.

\bibitem{Burgess:1993mg}
C.~Burgess, S.~Godfrey, H.~Konig, D.~London and I.~Maksymyk, \emph{{A Global
  fit to extended oblique parameters}},
  \href{https://doi.org/10.1016/0370-2693(94)91322-6}{\emph{Phys. Lett. B}
  {\bfseries 326} (1994) 276}
  [\href{https://arxiv.org/abs/hep-ph/9307337}{{\ttfamily hep-ph/9307337}}].

\bibitem{ABAZOV201188}
V.~Abazov, B.~Abbott, M.~Abolins, B.~Acharya, M.~Adams, T.~Adams et~al.,
  \emph{Search for a heavy neutral gauge boson in the dielectron channel with
  $5.4 ~fb^{-1}$ of $p\bar{p}$ collisions at $s=1.96~ tev$},
  \href{https://doi.org/https://doi.org/10.1016/j.physletb.2010.10.059}{\emph{Physics
  Letters B} {\bfseries 695} (2011) 88}.

\bibitem{Carena:2004xs}
M.~Carena, A.~Daleo, B.~A. Dobrescu and T.~M. Tait, \emph{{$Z^\prime$ gauge
  bosons at the Tevatron}},
  \href{https://doi.org/10.1103/PhysRevD.70.093009}{\emph{Phys. Rev. D}
  {\bfseries 70} (2004) 093009}
  [\href{https://arxiv.org/abs/hep-ph/0408098}{{\ttfamily hep-ph/0408098}}].

\bibitem{Accomando:2010fz}
E.~Accomando, A.~Belyaev, L.~Fedeli, S.~F. King and C.~Shepherd-Themistocleous,
  \emph{{Z' physics with early LHC data}},
  \href{https://doi.org/10.1103/PhysRevD.83.075012}{\emph{Phys. Rev. D}
  {\bfseries 83} (2011) 075012}
  [\href{https://arxiv.org/abs/1010.6058}{{\ttfamily 1010.6058}}].

\end{thebibliography}\endgroup
\end{document}